\documentclass[10pt]{article}

\usepackage{bm,amsmath,amsthm,amssymb}
\usepackage{amsfonts}
\usepackage{graphicx}
\usepackage{bm}
\usepackage{makeidx,epsfig,lscape}
\usepackage{color,colortbl}
\usepackage{fancyhdr}
\usepackage{xcolor,pict2e}

\renewcommand{\headrulewidth}{0pt}
\renewcommand{\footrulewidth}{0.5pt}

\definecolor{myaqua}{rgb}{0.0,0.5,0.55}
\definecolor{lightaqua}{rgb}{0.75,0.95,0.95}

\usepackage[colorlinks = true,
linkcolor = myaqua,
urlcolor  = blue,
citecolor = myaqua]{hyperref}

\usepackage{caption}
\usepackage{floatrow}

\def\lin#1#2{\textcolor[rgb]{0.6,0.6,0.6}{\vspace*{#1mm} \hrule
height 3 pt \vspace*{#2mm}}}
\def\bt{\begin{tabular}}
\def\et{\end{tabular}}
\def\and{\mbox{ and }}

\def\1{{\bf 1}}

\def\sectionn#1{\refstepcounter{section}{\color{myaqua}

\vskip 6mm

\noindent\Large\bf\thesection. #1}

\vskip 3mm}
\def\subsectionn#1{\refstepcounter{subsection}{\color{myaqua}

\vskip 5mm

\noindent\large\bf\thesubsection. #1}

\vskip 2mm}

\def\boxx#1#2#3#4#5{
{\linethickness{#4pt}\put(#1,#5){\color{myaqua}{\line(1,0){#3}}}}
\multiput(#1,#2)(0,#4){2}{\line(1,0){#3}}
\multiput(#1,#2)(#3,0){2}{\line(0,1){#4}}
}
\begin{document}

\fancyhead[L]{\hspace*{-13mm} \bt{l}{\bf Open Journal of Modern
Physics, 2016, 7, 662-679}\\ Published Online April 2016 in SciRes.
\href{http://www.scirp.org/journal/jmp}{\color{blue}{\underline{\smash{http://www.scirp.org/journal/jmp}}}}
\\
\href{http://dx.doi.org/10.4236/jmp.2016.77066}{\color{blue}{\underline{\smash{http://dx.doi.org/10.4236/jmp.2016.77066}}}}
\\ \et} \fancyhead[R]{\includegraphics{pic1.ps}}

$\mbox{ }$

\vskip 12mm

{ 

{\noindent{\huge\bf\color{myaqua}
On the Origin of Charge-Asymmetric Matter. II. Localized Dirac Waveforms}}
%
\\[6mm]
{\large\bf  Alexander Makhlin$^1$}}
\\[2mm]
{ 
$^1$ Rapid Research Inc, Southfield, MI, USA\\
Email: \href{mailto:amakhlin@comcast.net}{\color{blue}{\underline{\smash{amakhlin@comcast.net}}}}\\[1mm]

Received 25 February 2016; Accepted 25 April 2016; Published 28
April 2016
\\[4mm]
Copyright \copyright \ 2016 by author(s) and Scientific Research Publishing Inc. \\
This work is licensed under the Creative Commons Attribution International License (CC BY). \\
\href{http://creativecommons.org/licenses/by/4.0/}{\color{blue}{\underline{\smash{http://creativecommons.org/licenses/by/4.0/}}}}\\
\includegraphics{pic2.ps}

\lin{5}{7}

{ 
{\noindent{\large\bf\color{myaqua} Abstract}{\bf \\[3mm]
\textup{This paper continues the author's work \cite{PartI}, where a new
framework of the matter-induced physical geometry was built and an
intrinsic nonlinearity of the Dirac equation discovered. Here, the
nonlinear Dirac equation is solved and the localized
configurations are found analytically. Of the two possible types
of the potentially stationary localized configurations of the
Dirac field, only one is stable with respect to the action of an
external field and it corresponds to a positive charge. A
connection with the global charge asymmetry of matter in the Universe
and with the recently observed excess of the cosmic positrons is
discussed.
}}}\\[4mm]
{\noindent{\large\bf\color{myaqua} Keywords}{\bf \\[3mm]
Nonlinear Dirac field; Localization;  Cosmological charge asymmetry
}

\fancyfoot[L]{{\noindent{\color{myaqua}{\bf How to cite this
paper:}\color{black} Makhlin, A. (2016) On the Origin of
Charge-Asymmetric Matter. II. Localized Dirac Waveforms. Journal of
Modern Physics,{\bf 7}, 662-679}}
\href{http://dx.doi.org/10.4236/jmp.2016.77066}{\color{blue}
\underline{\smash{http://dx.doi.org/10.4236/jmp.2016.77066}}}}

\lin{3}{1}
\setcounter{page}{662}
\sectionn{Introduction}

{ \fontfamily{times}\selectfont
\noindent This paper continues the author's study of the long-standing
question of how the physical Dirac field of a real matter becomes
a finite-sized particle,  and it is approached here as a practical
problem.  The problem is posed and solved in a new framework of
the matter-induced affine geometry  \cite{PartI}, which deduces
the geometric relations in the space-time continuum from the
dynamic properties of the Dirac field. The intuitive argument of a
possible auto-localization of the Dirac field followed from an
observation \cite{PartI} that the local time flows slower at
higher invariant density, and then from the wave nature of the
Dirac equation. Its further consequence must be the (well-known
but not clearly understood) charge asymmetry of the observed
localized matter. In the present work, these qualitative
expectations are confirmed by explicit calculations.

The earlier developed \cite{PartI} mathematical background for the
present work is based on the following ideas and results. It is
observed that if at a point in spacetime continuum (the principal
differentiable manifold $\mathbb{M}$) a physical Dirac field is
defined, then the latter determines the tetrad of {\it Dirac
currents}. These are linearly independent and Lorentz-orthogonal
and can serve as local algebraic basis for any four-dimensional
vector space, including the infinitesimal displacements in
coordinate space.

The Dirac currents are employed as the Cartan's moving frame in
spacetime which, in its turn, results in the technique of
covariant derivatives for the vector and spinor fields. The
physics is naturally brought into this mathematical picture by the
equations of motion of the Dirac field, which made unnecessary an
artificial tangent (pseudo)Euclidean space. Differential
identities derived from equations of motion fully determine all
the components of the matter-induced affine connection (the Ricci
coefficients of rotation of the tetrad) in $\mathbb{M}$ and
without resorting to a particular coordinate system. Thus
determined connections completely define an {\it affine geometry}
(endowed with the connection but with no metric). Thus defined
connection depends on the Dirac field which makes the Dirac
equation nonlinear.

With known connections, it became possible to find the coordinate
lines and coordinate surfaces of the matter-induced affine
geometry, which have a clear physical meaning  and quite high
degree of symmetry. The congruence of lines of the timelike vector
current appeared to be normal, thus determining the family of the
hypersurfaces of the constant world time $\tau$. The lines of the
spacelike axial current appeared to be straight and their
congruence normal. They define the surfaces of the constant
distance $\rho$. The two-dimensional surfaces of constant $\rho$
at a given time $\tau$ were proved to be just spherical surfaces.

Below, the inevitable localization of the Dirac field into
particles observed in real world, but not explained by any theory
so far, is confirmed by the analytic solutions of the nonlinear
Dirac equation in one-body approximation. One of the solutions has
maximum near its center and is clearly associated with a stable
localized positive charge. Another one has minimum and is sought
to be an intrinsically unstable negative charge, which can be only
weakly localized by an external field.

The content of the paper is organized as follows. In
Sec.\ref{sec:Sec2} we use the previously developed \cite{PartI}
tools of the matter-induced affine geometry to write down the
Dirac equation in its most general coordinate-independent form.
Then, in Sec.\ref{sec:Sec3} we derive the formulae that connect
the Dirac matrices in the principal manifold $\mathbb{M}$ and in
arithmetic $\mathbb{R}^4$. In Sec.\ref{sec:Sec4}, the Dirac
equation in written down in a mixed representation, with
derivatives in $\mathbb{M}$, and coordinates and Dirac matrices in
$\mathbb{R}^4$. This representation is well suited for finding the
analytic solution. These solutions are found in Sec.\ref{sec:Sec5}
and their stability is discussed in Sec.\ref{sec:Sec7}.   The
conceptual questions of the charge-asymmetric real world are
briefly discussed in the Summary.

\renewcommand{\headrulewidth}{0.5pt}
\renewcommand{\footrulewidth}{0pt}

\pagestyle{fancy}
\fancyfoot{}
\fancyhead{} 
\fancyhf{}
\fancyhead[RO]{\leavevmode \put(-90,0){\color{myaqua}Alexander Makhlin} \boxx{15}{-10}{10}{50}{15} }
\fancyhead[LE]{\leavevmode \put(0,0){\color{myaqua}Alexander Makhlin}  \boxx{-45}{-10}{10}{50}{15} }
\fancyfoot[C]{\leavevmode
\put(0,0){\color{lightaqua}\circle*{34}}
\put(0,0){\color{myaqua}\circle{34}}
\put(-2.5,-3){\color{myaqua}\thepage}}

\renewcommand{\headrule}{\hbox to\headwidth{\color{myaqua}\leaders\hrule height \headrulewidth\hfill}}
\sectionn{The framework. \label{sec:Sec2}}
\renewcommand{\theequation}{2.\arabic{equation}}
\setcounter{equation}{0}
{ \fontfamily{times}\selectfont
\noindent

In the first part of this work we explored differential identities
for the four Dirac currents, vector current $\bm j$, axial current
$\bm J$, and two \textquotedblleft charged currents\textquotedblright, $\bm \Theta$
and $\bm \Phi$. Using them, we found all components of the affine
connection $\omega_{ABC}$, as well as connection $\Gamma_B$ of the
Dirac field in principal manifold $\mathbb{M}$,
\begin{eqnarray}\label{eq:E2.1}
\Gamma_B = ieA_B+(1/4)\omega_{ACB}\rho_1\alpha^A\rho_1\alpha^C.
\end{eqnarray}
The connection (\ref{eq:E2.1}) determines the covariant derivative
of the Dirac field and it enters the Dirac equation as
$\alpha^B\Gamma_B$,
\begin{equation}\label{eq:E2.2}
\alpha^B[\partial_B\psi -\Gamma_B\psi]=-im\rho_1\psi.
\end{equation}
The nonzero elements of the $\omega_{ABC}$ in the tetrad
basis of the normalized Dirac currents ${\bm e}_A$ are as follows,
\begin{eqnarray}\label{eq:E2.3}
\omega_{030}=-\omega_{131}=-\omega_{232}=Q,~~~\omega_{12D}=2e{\tilde
A}_{[D]}, ~~~(D=0,1,2,3),
\end{eqnarray}
where $Q\equiv\partial_{[3]}\ln{\cal R}=-m{\cal P}/{\cal R}=
-m\sin{\cal Y}$ is the derivative of the invariant density ${\cal
R}$ in the direction of the axial current and it has an algebraic
representation via the pseudoscalar density ${\cal P}$. These
formulae assume that ${\tilde A}_{[D]}=+A_D$ for the right-handed
spatial triad $\bm{e}_{[1]},\bm{e}_{[2]},\bm{e}_{[3]}$ with ${\bm
\Theta}={\cal R}\bm{e}_{[1]}$, ${\bm \Phi}={\cal R}\bm{e}_{[2]}$ and
the naturally outward directed axial current ${\bm J}={\cal R}{\bm
e_3}$, i.e. $[{\vec e}_{[1]}\times {\vec e}_{[2]}]={\vec e}_{[3]}$
[c.f Eqs. (\ref{eq:A.4}), (\ref{eq:A.5})]. When the latter is
directed inward, but we still wish ${\bm e_{[3]}}$ to point outward,
then we have to take ${\bm \Theta}={\cal R}{\bm e_{[2]}}$, ${\bm
\Phi}={\cal R}{\bm e_{[1]}}$ and replace $\omega_{12D} \to
\omega_{21D}=-\omega_{12D}$ (or ${\tilde A}_{[D]}=-A_D$) in
Eqs.(\ref{eq:E2.3}) \footnote{Throughout this paper, when uppercase
index $A$ of the basis ${\bm    e}_A \equiv {\bm e}_{[A]} $,
$(A=0,1,2,3)$ takes a particular numeric value we put it in
brackets, $[0],[1],...$. The lowercase indices $a$ that are related
to the tetrad ${\bm h}_a\equiv{\bm h}_{(a)}$ are put in parentheses,
$(0),(1),...$.}.

It is instructive to see how the operator $D_A =\partial_A-
\Gamma_A$ carries out the parallel transport of the Dirac spinor
$\psi$ in different directions. Substituting the results
(\ref{eq:E2.3}) into connection (\ref{eq:E2.1}), it is
straightforward to obtain,
\begin{eqnarray}\label{eq:E2.4}
\alpha^{[0]}\Gamma_0={1\over2}Q\alpha^{[3]}+2ieA_{[0]}\alpha^{[0]}\Pi
={1\over2}Q\alpha^{[3]}+ie[A_{[0]}\alpha^{[0]}- {\tilde A}_{[0]}
\rho_{[3]} \alpha^{[3]}],~~~\nonumber\\
\alpha^{[3]}\Gamma_3=+2ieA_{[3]}\alpha^{[3]}\Pi=ie[A_{[3]}\alpha^{[3]}-
{\tilde A}_{[3]}\rho_{[3]} \alpha^{[0]}]  ,~~~\nonumber\\
\alpha^{[1]}\Gamma_1={1\over2}Q\alpha^{[3]}+2ieA_{[1]}\alpha^{[1]}\Pi
={1\over2}Q\alpha^{[3]}+ie[A_{[1]}\alpha^{[1]}+ i{\tilde
A}_{[1]}\alpha^{[2]}],~~~\nonumber\\
\alpha^{[2]}\Gamma_2={1\over2}Q\alpha^{[3]} +2ieA_{[2]}
\alpha^{[2]}\Pi= {1\over2}Q\alpha^{[3]} +ie[A_{[2]} \alpha^{[2]} -
i{\tilde A}_{[2]} \alpha^{[1]}],~~~
\end{eqnarray}
where $\Pi=(1\mp i\gamma^{[1]}\gamma^{[2]})/2=S^{-1}(1\pm
\sigma^{(3)})S /2$. The upper and lower signs in the projector $\Pi$
(accordingly, the sign in ${\tilde A}_{[D]}=\pm A_{[D]}$)
correspond to the outward and inward directions of the axial
current, respectively, which then determines the right- and left-
oriented spatial triplets ${\bm e}_{[1]},{\bm e}_{[2]},{\bm e}_{[3]}$. It will
be shown below, that, from the perspective of the localized
solutions, this orientation is translated into the bump of the
positive charge and to the dip of the negative one, respectively,
i.e. $\pm=-{\rm sign} (\partial_{[3]}{\cal R})$. Therefore,
depending on this sign, only the {\it locally inward or locally
outward} components, $(d_L,d_R)$ or $(u_L,u_R)$, interact with the
electromagnetic potential but with the doubled coupling constant
$2e$. In a sense, the charge conjugation goes together with
spatial reflection. The matrix $\rho_3$ differentiate between the
right and left components.

With the connection (\ref{eq:E2.4}) the Dirac equation becomes a
nonlinear equation and its explicit form reads as,
\begin{eqnarray}\label{eq:E2.5}
\alpha^{[0]} \bigg[\partial_{[0]} - ieA_{[0]}+ i \rho_{[3]} e
\tilde{A}_{[3]}\bigg]\psi + \alpha^{[3]} \bigg[\partial_{[3]} -
ieA_{[3]}+ i \rho_{[3]} e\tilde{A}_{[0]} -(3/2)Q \bigg]\psi
\nonumber\\* + \alpha^{[1]} \bigg[\partial_{[1]} - ieA_{[1]}-
e\tilde{A}_{[2]}\bigg]\psi + \alpha^{[2]} \bigg[\partial_{[2]}-
ieA_{[2]}+ e\tilde{A}_{[1]} \bigg]\psi +im\rho_{[1]}\psi=0,~~~
\end{eqnarray}
where anomalous term $-3Q/2$ singles out the direction of the
axial current among others even when $A_\mu=0$.

This equation is valid in every connected domain where ${\cal
R}^2>0$ and the Dirac currents define a non-degenerate orthogonal
tetrad $e^\mu_{A}(\psi)$. As anticipated, it is invariant in a
most broad sense -- it depends neither on choice of coordinates
$x^\mu$ in $\mathbb{R}^4$ nor on a tetrad system $h_a^\mu$ (also
in $\mathbb{R}^4$) not even on a particular choice of the
$\gamma$-matrices. The latter is always taken for granted since
one can introduce a new Dirac field $\psi^\prime=S\psi$ leaving
the gamma matrices unchanged. But this trick works only for
re-parameterizations in $\mathbb{R}^4$, i.e. change of the Lorentz
frame or transformations between orthogonal coordinates. It cannot
be employed in the principal manifold $\mathbb{M}$ just because
the Dirac field is a coordinate scalar.

Finally, Eq.(\ref{eq:E2.5}) is nonlinear because both the
connection $\omega_{ACB}$ and the Dirac matrices $\alpha^A=
V^A_a(\psi)\alpha^a$ in it depend on the Dirac field $\psi\in
\mathbb{M}$. The dependence of $\omega_{ACB}$ on the Dirac field
is due to (\ref{eq:E2.3}). The dependence of the Dirac matrices on
$\psi$, $\alpha^A=V^A_a(\psi)\alpha^a$, is not so explicit but not
less important and it cannot be avoided. Indeed, in the basis
$[A]$ each of the currents ${\bm J}_A$ has only one nonzero
component, e.g.,
\begin{eqnarray}
j^A=\psi^+\alpha^A \psi=V^A_a j^a={\cal R}V^A_a V^a_0={\cal R}
\delta^A_{[0]}.\nonumber
\end{eqnarray}
The latter cannot be achieved without an explicit dependence
$\alpha^A(\psi)$. Indeed, with $\psi\in \mathbb{M}$ and numerical
matrices $\alpha^a$ the current $j^a$ will have all components.
Obviously, this is a significant technical difficulty. However,
only this dependence solves a conceptual problem of independence
of the equation of motion for the physical Dirac field in
$\mathbb{M}$ on a particular choice of the tetrad ${\bm h}_a$ and
of the matrices $\alpha^a$ in tangent $T_p$. Therefore, we begin
with the establishing rules of transformation of the 16 Dirac
matrices between $\mathbb{M}$ and $\mathbb{R}^4$.
\newpage
\sectionn{Dirac matrices in principal manifold $\mathbb{M}$}\label{sec:Sec3}
\renewcommand{\theequation}{3.\arabic{equation}}
\setcounter{equation}{0}
{ \fontfamily{times}\selectfont
\noindent

Historically, the Dirac equation for the free field $\psi$ was
formulated as $i\alpha^a\partial_a\psi -m\beta\psi=0$ with the aid
of {\it Hermitian} Dirac matrices $\alpha^a=(\alpha^a)^+$ and
$\beta=\beta^+$, which satisfy the commutation relations,
\begin{equation}\label{eq:E3.1}
\alpha^a \beta \alpha^b  +  \alpha^b \beta \alpha^a= 2\beta
\eta^{ab}~,~~~~\alpha^a \beta  + \beta \alpha^a=0,~~~ \beta^2=1.
\end{equation}
Usually one assumes that $\alpha^a= (1,\alpha^i)$; $a=0,1,2,3$;
$i=1,2,3$ (so that $\alpha^0=1$ is a unit matrix) but this is not
required. An apparently symmetric form of commutation relations
(\ref{eq:E3.1}) emerges (along with the equation,
$i\gamma^a\partial_a\psi -m\psi=0$) in terms of the matrices
$\gamma^a=(\gamma^{(0)},\gamma^i)= (\beta,\beta\alpha^i)$,
\begin{eqnarray}\label{eq:E3.2}
\gamma^a\gamma^b+ \gamma^b\gamma^a=2\eta^{ab}.
\end{eqnarray}
Neither of these matrices is uniquely defined. However, if there
exist two sets of the matrices, $\gamma^a$ and $\gamma^{[A]}$,
that satisfy (\ref{eq:E3.2}) then, according to the Pauli's
fundamental theorem, there exists such a nonsingular $S$, that
\begin{eqnarray}\label{eq:E3.3}
\gamma^{[\kappa]}=S^{-1}\gamma^{(\kappa)} S,
\end{eqnarray}
where $\kappa=0,1,2,3$, $[\kappa]$ is a number standing for
superscript $A$ and $(\kappa)$ is the same number for superscript
$a$. There are sixteen linearly independent $4\times 4$ matrices
$O_p=(1,\gamma^a,\gamma^a \gamma^b,...)$, all of which are the
products of 1,2,3 or 4 different gamma.  Therefore, $O_{[p]}=
S^{-1}O_p S=(1,\gamma^{[a]},\gamma^{[a]} \gamma^{[b]},...)$, which
is an indisputable technical advantage.

By their definition, the matrices $\gamma^a$ are not Hermitian.
However, since $\beta$ and $\alpha^i$ are Hermitian and
anti-commuting, the Hermit-conjugated $\gamma$-matrices are
$(\gamma^a)^+=\gamma^{(0)}\gamma^a\gamma^{(0)}$. If, by the same token,
$\gamma^{[A]}=\gamma^{[0]}\alpha^{[A]}$ (with Hermitian
$\gamma^{[0]}$ and $\alpha^{[A]}$), then
$(\gamma^{[A]})^+=\gamma^{[0]}\gamma^{[A]}\gamma^{[0]}$, which
yields, $$ S^{-1}\gamma^{(A)} S=
\gamma^{[A]}=\gamma^{[0]}(\gamma^{[A]})^+\gamma^{[0]}=
\gamma^{[0]}(S^{-1}\gamma^{A}S)^+\gamma^{[0]}=
(\gamma^{[0]}S^+\gamma^{(0)})\gamma^A
(\gamma^{[0]}S^+\gamma^{(0)})^{-1}.$$ Multiplying this  by $S$ from
the left and by $\gamma^{[0]}S^+\gamma^{(0)}$ from the right, we
find,
\begin{eqnarray}\label{eq:E3.4}
\gamma^A(S\gamma^{[0]}S^+\gamma^{(0)})=(S\gamma^{[0]}S^+\gamma^{(0)})\gamma^A.
\end{eqnarray}
The matrix $(S\gamma^{[0]}S^+\gamma^{(0)})$ commutes with all the
matrices $\gamma^A$ and must be the unit matrix, {\it viz.},
\begin{eqnarray}\label{eq:E3.5}
\gamma^{[0]}S^{-1}= S^+\gamma^{(0)}.
\end{eqnarray}
On the one hand, we can continue as
\begin{eqnarray}\label{eq:E3.6}
\alpha^{[A]}= \gamma^{[0]}\gamma^{[A]} = \gamma^{[0]}
S^{-1}\gamma^A S=S^+\gamma^{(0)}\gamma^A S=S^+\alpha^A S.
\end{eqnarray}
On the other hand, condition (\ref{eq:E3.5}) means that
$\gamma^{[0]}= S^+\gamma^{(0)}S \neq S^{-1}\gamma^{(0)} S $, which
conflicts with Eq.(\ref{eq:E3.3}), because matrix $S$ is not
unitary. This conflict can be avoided by adopting a slightly
different agreement (that does not affect any of the common usages
of the gamma-matrices). Namely, let us denote $\beta=\rho_1$ and
define $\gamma$-matrices as $\gamma^a=\rho_1\alpha^a$ and
$\gamma^{[A]}=\rho_1\alpha^{[A]}$. Now we must replace  both
$\gamma^{[0]}$ and $\gamma^{(0)}$ in Eq.(\ref{eq:E3.4}) by $\rho_1$,
so that $S^+\rho_1=\rho_1 S^{-1}$ and $$\gamma^{[a]}=\rho_1
\alpha^{[a]}=\rho_1 S^+\alpha^{(a)} S =S^{-1}\rho_1\alpha^{(a)} S=
S^{-1}\gamma^{(a)} S, ~~a=0,1,2,3,$$ in compliance with
(\ref{eq:E3.3}). Choosing $\alpha^{(0)}=1$, we have $\gamma^{(0)}=\rho_1$,
$\alpha^{[0]}=S^+S$, $\gamma^{[0]}=\rho_1 S^+ S= S^+
S\rho_1=(\gamma^{[0]})^+$.

Throughout this paper, we are only interested  in a special case of
the transformations (\ref{eq:E3.3}) and (\ref{eq:E3.6}),
\begin{eqnarray}\label{eq:E3.7}
\gamma^{[A]}=V^A_a\gamma^a,~~~\alpha^{[A]}=V^A_a\alpha^a,
\end{eqnarray}
where the transformation matrix $V^A_a$ is real and has the
properties,
\begin{eqnarray}\label{eq:E3.8}
V^a_A V_a^B=\delta_A^B, ~~~ V^a_A V_b^A=\delta_b^a.~~
\end{eqnarray}
Then the commutation relations (\ref{eq:E3.1}) are the same  for
$\gamma^a$ and $\gamma^A$ and $S$ must be a solution of the matrix
equation,
\begin{eqnarray}\label{eq:E3.9}
\alpha^{[A]}=S^+\alpha^A S=V^A_a(\psi)\alpha^a.
\end{eqnarray}
Though $V^A_a$ has a character of a Lorentz transformation, it has
no infinitesimal prototype. Since $ S^+=\rho_1 S^{-1}\rho_1$, we
also have a habitual $\gamma^{[A]}=\rho_1 \alpha^{[A]}=
S^{-1}\gamma^A S= V^A_a(\psi)\gamma^a$. However, in the basis of
matrices $\gamma^{[A]}$, the Pauli-conjugated Dirac spinor must be
defined as $\overline{\psi}= \psi^+\rho_1$ and {\it not} as
$\overline{\psi}=\psi^+\gamma^{[0]}$.

The set $O_p$ of 16 linearly independent elements of Clifford
algebra comprised of various products of the $\gamma^a$- (or  the
$\gamma^A$-) matrices is in one-to-one correspondence with 16
Hermitian matrices, $(1,\rho_i,\sigma^i, \rho_i\sigma^k
=\sigma^k\rho_i)$, $i,k=1,2,3$, where $\rho_1=\gamma^0$,
$\rho_2=\gamma^1\gamma^2\gamma^3$, $\rho_3=i\gamma^0
\gamma^1\gamma^2\gamma^3=i\rho_1\rho_2$ and $\sigma^i=
i\rho_2\gamma^i=i\gamma^1\gamma^2\gamma^3\gamma^i=
\rho_3\alpha^i$. The Dirac matrices,  $~\rho_i~$ and $~\sigma^i~$,
satisfy the same commutation relations as the Pauli matrices,
$\sigma^i\sigma^k=\delta_{ik}+i\epsilon_{ikl}\sigma^l$, and
$\rho_a\rho_b =\delta_{ab}+i\epsilon_{abc}\rho_c$.  Finally, it is
straightforward to check that the matrix $\rho_3=i\gamma^0\gamma^1
\gamma^2\gamma^3$ (commonly known as $-\gamma^5$) is an invariant
of transformations (\ref{eq:E3.3}),
\begin{eqnarray}\label{eq:E3.10}
\rho_{[3]}={i\over 4!}\epsilon_{ABCD}\gamma^A\gamma^B
\gamma^C\gamma^D={i\over 4!}\epsilon_{ABCD}V^A_aV^B_bV^C_cV^D_d
\gamma^a\gamma^b \gamma^c\gamma^d={i\over 4!}
\epsilon_{abcd}\gamma^a\gamma^b \gamma^c\gamma^d=\rho_3
\end{eqnarray}
Then the matrix $\rho_2=i\rho_1\rho_3$ is transformed like
$\rho_1$, so that
\begin{eqnarray}\label{eq:E3.11}
\rho_{[3]}=S^{-1}\rho_3 S=\rho_3,~~ \rho_{[i]}=S^+\rho_i S=\rho_i,
~~~i=1,2.
\end{eqnarray}
As long as $S^+\rho_3S=S^+S\rho_3=\rho_3S^+S= \rho_3\alpha^{[0]}$,
the matrices $\sigma$ on the $\mathbb{M}$, being defined as
$\sigma^{[I]}=\rho_3\alpha^{[I]}$, are transformed as
\begin{eqnarray}\label{eq:E3.12}
\sigma^{[I]}=S^+\sigma^I S=\rho_3(V_{(0)}^I\alpha^{(0)}+ V^I_{j}\alpha^{j}) =
V_{(0)}^I\rho_3+ V^I_j\sigma^j
\end{eqnarray}
(as it should be for the spatial components of the axial current
${\cal J}^a$) \footnote{Then the charge-conjugated spinor
$\psi_{\bm c}={\bm C}\psi^* =\rho_2\sigma^2 \psi^*$ becomes
$\psi_{\bm c}=\rho_2\sigma^{[2]} \psi^*$. In particular,
$\Lambda_{(-)}^a= \psi^+\alpha^a\psi_{\bm c} \to
\Lambda_{(-)}^{[a]}= \psi^+\alpha^{[a]}S^{-1}\rho_2 \sigma^2 S
\psi^*=\psi^+ S^+ \alpha^{a}\rho_2\sigma^{2} S \psi^*$. At the
same time, $\gamma^{[0]}\gamma^{[I]}=S^{-1}\alpha^I S $ and $i
\gamma^{[1]}\gamma^{[2]}=S^{-1}\sigma^3 S$,~...}.

\sectionn{The nonlinear Dirac equation, explicitly  \label{sec:Sec4}}
\renewcommand{\theequation}{4.\arabic{equation}}
\setcounter{equation}{0}
{ \fontfamily{times}\selectfont
\noindent

So far, we have been studying the general geometric properties of the
Dirac field in the scope of the affine geometry and carefully
avoiding any assumptions about what {\it a solution of the Dirac
equation} that has these properties can be. All the previously
established \cite{PartI} properties of the Dirac currents belong
(along with the Dirac field itself) to the principal
differentiable manifold $\mathbb{M}$. Without resorting to any
particular coordinate  manifold $\mathbb{R}^4$  we have
established in \cite{PartI} the following facts:\\ {\bf(i)} The
congruence of lines of the vector field $e_{[0]}^\mu$ is normal.
The family $S_{(123)}$ of hypersurfaces, $\tau(x)={\rm const}$, of
the constant world time $\tau$ is extrinsically flat; $\tau$ is a
holonomic coordinate and it can be taken for $x^0$ in
$\mathbb{R}^4$.\\ {\bf(ii)} The congruence of lines of the vector field
$e_{[3]}^\mu$ is normal and geodesic. The hypersurfaces
$S_{(012)}$ of the constant radius $\rho$ have constant extrinsic
curvature and the holonomic coordinate $\rho$ can serve as $x^3$
in $\mathbb{R}^4$. \\{\bf(iii)} The two-dimensional surfaces $S_{(12)}$
of constant $\tau$ and $\rho$ are just spheres, i.e. umbilical
(with two equal Gauss'curvatures) surfaces with constant mean
(extrinsic) curvature $H=m{\cal P/R}=-m\partial_{[3]}\ln{\cal R}$.
The latter is determined by the Dirac field within principal
manifold $\mathbb{M}$ and depends only on the radius $\rho$. The
intrinsic (sectional) curvature, $R^t_{1212}=2e(\partial_{[1]}
A_{[2]}-\partial_{[2]} A_{[1]})- 4e^2(A_{[1]}^2+ A_{[2]}^2)
=2eF_{12}=2eB_{[3]}$, is due to the {\it external} electromagnetic
field. It coincides with projection of the magnetic field onto the
direction of the axial current.
\\{\bf(iv)} The two-dimensional surfaces $S_{(03)}$ are covered by a
well-defined coordinate net formed by the streamlines of the
vector and axial currents. This net can be identically mapped
between the principal manifold $\mathbb{M}$ and the arithmetic
$\mathbb{R}^4$.

These general observations can be summarized as follows. For any
solution of the Dirac equation, which is not homogeneous in
spatial directions, {\it the spherical symmetry is the property of
a solution, thus being a dynamic symmetry}.

In order to find a solution of the Dirac equation, one has to
specify a coordinate basis in $\mathbb{R}^4$ and a basis of the
Dirac matrices. Here, we shall employ the numerical matrices
$\alpha^a$ in spinor representation (\ref{eq:A.2}) and associate
them with a tetrad ${\bm h}^\mu_{(a)}$. Then, $\alpha^A=
V^A_a\alpha^a$, while the derivatives ${\cal D}_{[A]}$ will stay
in the basis ${\bm e}_A$, which is associated with coordinate surfaces determined in the principal manifold $\mathbb{M}$. In this mixed representation, Dirac
equation reads as
\begin{eqnarray}\label{eq:E4.1}
\left[ \begin{array}{c c c c}  V_{(0)}^A +V_{(3)}^A & V_{(1)}^A
-iV_{(2)}^A & 0 & 0 \\ V_{(1)}^A +iV_{(2)}^A & V_{(0)}^A
-V_{(3)}^A & 0 & 0           \\ 0 & 0 & V_{(0)}^A -V_{(3)}^A &
-V_{(1)}^A +iV_{(2)}^A          \\ 0 & 0 & -V_{(1)}^A - iV_{(2)}^A
&V_{(0)}^A +V_{(3)}^A \end{array}\right]
\!\!\left[ \begin{array}{c}{\cal D}_A (u_R e^{i\phi_R^u})
\\{\cal D}_A (d_R e^{i\phi_R^d})\\{\cal D}_A (u_L e^{i\phi_L^u})
\\{\cal D}_A(d_L e^{i\phi_L^d})\end{array}\right]\!\!= -im
\!\!\left[ \begin{array}{c} u_L e^{i\phi_L^u}
\\d_L e^{i\phi_L^d}\\u_R e^{i\phi_R^u}
\\d_R e^{i\phi_R^d}\end{array}\right].~~~~~
\end{eqnarray}
The operators ${\cal D}_A$, which are copied from
Eq.(\ref{eq:E2.5}), are as follows,
\begin{eqnarray}\label{eq:E4.2}
{\cal D}_{[0]}=\partial_{[0]}-ieA_{[0]}+ i\rho_3
e\tilde{A}_{[3]},~~~~~~~~~~~~~~~  {\cal D}_{[1]}=
\partial_{[1]} -ieA_{[1]}- e\tilde{A}_{[2]},~\nonumber\\*
{\cal D}_{[3]}=\partial_{[3]}-ieA_{[3]}+ i\rho_3
e\tilde{A}_{[0]}-3Q/2,~~~~~{\cal D}_{[2]}=\partial_{[2]}
-ieA_{[2]}+ e\tilde{A}_{[1]},~
\end{eqnarray}
where $\rho_3$ differentiate between the right and left components
and it stands for $+1$ for $u_R,d_R$ and for $-1$ for $u_L,d_L.$
The coordinate net formed by the integral lines of the tetrad vectors
${\bm e}_{[0]}$ and ${\bm e}_{[3]}$ that covers the two-dimensional
surface $S_{(03)}$ in $\mathbb{M}$ is holonomic and
the vectors ${\bm h}_{(0)},{\bm h}_{(3)}$ in $\mathbb{R}^4$ can be
chosen tangent to this surface. In order for the other two tetrad
vectors, ${\bm h}_{(1)}$ and ${\bm h}_{(2)}$, to be normal to this
surface, it is necessary that the components $V_{[0]}^{(1)}=
V_{[0]}^{(2)} =V_{[3]}^{(1)} =V_{[3]}^{(2)}=0$. Just by inspection
of Eqs.(\ref{eq:A.4}), we see that this is possible only when
either $d_R=d_L=0$ or $u_R=u_L=0$. In both cases, as seen from
Eqs.(\ref{eq:A.5}), we have $V_{[1]}^{(0)}= V_{[1]}^{(3)}=
V_{[2]}^{(0)}=V_{[2]}^{(3)}=0$. In other words, the spacetime with
the matter-induced anholonomic basis can be viewed as a direct
product of the two-dimensional subspaces, $S_{(03)}\bigotimes
S_{(12)}$.  This is sufficient to treat the up- and
down-polarizations separately,
\begin{eqnarray}\label{eq:E4.3}
\psi_u=\left[ \begin{array}{c} u_R \exp({i\phi_R^u })
\\0\\u_L \exp({i\phi_L^u})
\\0\end{array}\right],~~~\psi_d=\left[ \begin{array}{c} 0 \\
d_R \exp({i\phi_R^d }) \\0\\d_L \exp({i\phi_L^d})
\end{array}\right].
\end{eqnarray}

Having only $u_R,u_L$ or $d_R,d_L$ components, the states $\psi_u$ and $\psi_d$ cannot bear quantum numbers of an angular momentum. For the up-polarized $\psi_u$, we have ${\cal J}^{(3)}=+|{\cal J}^{(3)}|$, $Q\equiv\partial_{[3]}\ln{\cal R}=-m\sin{\cal Y}<0$ . In this case
[C.f. (\ref{eq:A.4})-(\ref{eq:A.6})], ${\cal R}={\cal R}_u
=2u_Ru_L$ and the matrix $\alpha^{(a)} V_{(a)}^{[A]}$ in the
l.h.s. of Eq.(\ref{eq:E4.1}) simplifies to  $$ V_{(0)}^{[0]}
+V_{(3)}^{[0]}= V_{(0)}^{[3]} +V_{(3)}^{[3]} =u_R/u_L,
~V_{(0)}^{[0]} -V_{(3)}^{[0]}=V_{(3)}^{[3]} -V_{(0)}^{[3]}
=u_L/u_R,~ V_{(1)}^{[1]}\pm iV_{(2)}^{[1]}= \mp i (V_{(1)}^{[2]}
\pm iV_{(2)}^{[2]})=e^{\mp i(\phi_L^u+\phi_R^u)}.$$ Accordingly,
system (\ref{eq:E4.1}) for $\psi_u$ becomes
\begin{eqnarray}\label{eq:E4.4}
u_R[{\cal D}_{[0]}+{\cal D}_{[3]}]u_Re^{i\phi_R^u} = -im
u_L^2e^{i\phi_L^u},~~~e^{-i(\phi_R^u+\phi_L^u)} [{\cal
D}_{[1]}+i{\cal D}_{[2]}] u_Re^{i\phi_R^u}=0, \nonumber\\
u_L[{\cal D}_{[0]}-{\cal D}_{[3]}] u_Le^{i\phi_L^u}
=-imu_R^2e^{i\phi_R^u},~~~ e^{-i(\phi_R^u+\phi_L^u)} [{\cal
D}_{[1]} +i{\cal D}_{[2]}] u_Le^{i\phi_L^u}=0.
\end{eqnarray}

For the down-polarized $\psi_d$, we have ${\cal J}^{(3)}=-|{\cal J}^{(3)}|$, $Q=\partial_{[3]}\ln{\cal R}=+m\sin{\cal Y}>0$. Here, ${\cal
R}={\cal R}_d =2d_Rd_L$ and the elements of the matrix in the
l.h.s. of Eq.(\ref{eq:E4.1}) become, $$ V_{(0)}^{[0]}
-V_{(3)}^{[0]}= V_{(0)}^{[3]} -V_{(3)}^{[3]})
=d_R/d_L,~V_{(0)}^{[0]} +V_{(3)}^{[0]} =-(V_{(0)}^{[3]}
+V_{(3)}^{[3]}) =d_L/d_R,~ -(V_{(1)}^{[1]}\pm iV_{(2)}^{[1]})= \mp
i (V_{(1)}^{[2]}\pm iV_{(2)}^{[2]})=e^{\mp
i(\phi_L^d+\phi_R^d)}.$$ Now, the system (\ref{eq:E4.1}) reads as
\begin{eqnarray}\label{eq:E4.5}
d_R[{\cal D}_{[0]}+{\cal D}_{[3]}]d_Re^{i\phi_R^d} = -im
d_L^2e^{i\phi_L^d},~~~ e^{-i(\phi_R^d+\phi_L^d)}[{\cal
D}_{[1]}+i{\cal D}_{[2]}] d_Re^{i\phi_R^d}=0, \nonumber\\
d_L[{\cal D}_{[0]}-{\cal D}_{[3]}] d_Le^{i\phi_L^d}
=-imd_R^2e^{i\phi_R^d},~~~ e^{-i(\phi_R^d+\phi_L^d)} [{\cal
D}_{[1]}+i{\cal D}_{[2]}] d_Le^{i\phi_L^d}=0.
\end{eqnarray}
Remembering about the sign due to $\rho_3$, we obtain the
following formulae for all the differential operators involved,
\begin{eqnarray}\label{eq:E4.6}
{\cal D}_{[0]}+ {\cal D}_{[3]}=\partial_{[0]}+ (\partial_{[3]}
-(3/2)Q)-ie[(A_{[0]}-\tilde{A}_{[0]})+(A_{[3]}-
\tilde{A}_{[3]})],\nonumber\\*{\cal D}_{[0]}- {\cal D}_{[3]}
=\partial_{[0]}- (\partial_{[3]} -{3\over 2}Q)-
ie[(A_{[0]}-\tilde{A}_{[0]})-(A_{[3]}-\tilde{A}_{[3]})],\nonumber
\\* {\cal D}_{[1]}+ i{\cal D}_{[2]}=\partial_{[1]}+
i\partial_{[2]} -ie[(A_{[1]}-\tilde{A}_{[1]})
+i(A_{[2]}-\tilde{A}_{[2]})],\nonumber\\* {\cal D}_{[1]}- i{\cal
D}_{[2]}=\partial_{[1]}- i\partial_{[2]}
-ie[(A_{[1]}+\tilde{A}_{[1]})-i(A_{[2]}+\tilde{A}_{[2]})].
\end{eqnarray}
In Eqs.(\ref{eq:E4.4}) and (\ref{eq:E4.5}), the operator ${\cal
D}_{[0]}+ {\cal D}_{[3]}$ acts only on $u_R$ and $d_R$ while
${\cal D}_{[0]}- {\cal D}_{[3]}$ only on $u_L$ and $d_L$.

\sectionn{Solutions of the nonlinear equations  \label{sec:Sec5}}
\renewcommand{\theequation}{5.\arabic{equation}}
\setcounter{equation}{0}
{ \fontfamily{times}\selectfont
\noindent
So far we were expanding the vector of spacetime displacement
$dx^\mu$ in terms of the basis ${\bm e}_A$ of the tetrad
determined by the Dirac currents $dx^\mu= e^\mu_A dS^A$. But the
true physical variables are the world time $\tau$ and the distance
$\rho$. They are holonomic coordinates, because $d\tau={\cal R}dS^{[0]}$
and $d\rho={\cal R}dS^{[3]}$ are the total differentials of the
independent coordinates $dx^\mu\in \mathbb{R}^4$,
\begin{eqnarray}\label{eq:E5.1}
\tau_2-\tau_1= \int_{x(\tau_1)}^{x(\tau_2)} j_\mu(x) dx^\mu =\int
{\cal R}dS^{[0]},~~\rho_2-\rho_1=
\int_{x(\rho_1)}^{x(\rho_2)}{\cal J}_\mu(x) dx^\mu =\int \pm {\cal
R}dS^{[3]}.
\end{eqnarray}
Here, the upper sign is for the $\psi_u$, where ${\cal J}_3=
u_R^2+u_L^2>0$. The lower sign is for $\psi_d$, where ${\cal J}_3=
-d_R^2-d_L^2<0$ and the axial current is directed inward. The
world time $\tau$ and the radial variable $\rho$, being defined as
invariants in $\mathbb{M}$, can immediately be used in arithmetic
$\mathbb{R}^4$.

\subsectionn{Reduction to the physical variables}

At the points where $j^{(3)}=V^{(3)}_{[0]}=0$ and ${\cal J}^{(0)}=
V^{(0)}_{[3]}=0$ (in general, a 2-d surface) the relation between
spatial components, $[{\vec \Theta} \times{\vec \Phi}]/{\cal R}^2
= +{\vec {\cal J}}/{\cal R}$, is an algebraic identity. For the
axial current directed outward, i.e. ${\cal J}_3>0$, we take ${\cal
J}^\mu=+{\cal R}e_{[3]}^\mu$, $\Theta^\mu={\cal R} e_{[1]}^\mu$
and $\Phi^\mu={\cal R} e_{[2]}^\mu$,  so that ${\vec e}_{[3]}=
[{\vec e}_{[1]}\times {\vec e}_{[2]}]$. In this case, we change
the variables in Eq.(\ref{eq:E4.4}) as follows,
\begin{eqnarray}\label{eq:E5.2}
\partial_{[0]}\to m{\cal R}\partial_{\tau}, ~\partial_{[3]}\to
m{\cal R}\partial_{\rho},~eA_{[0]}\to{e\over m}{\cal R}A_\tau,~
eA_{[3]}\to{e\over m}{\cal R}A_\rho,~\nonumber\\{\omega_{120}\over
2}= e{\tilde A}_{[0]}\to{e\over m}{\cal R}{\tilde A}_\tau,~
{\omega_{123}\over 2}= e{\tilde A}_{[3]}\to{e\over m}{\cal R}
{\tilde A}_\rho,~e{\tilde A}_{[1]}\to {e\over m}{\tilde
A}_{[1]},~~e{\tilde A}_{[2]}\to {e\over m}{\tilde A}_{[2]}
\end{eqnarray}
Adopting the physical variables (\ref{eq:E5.2}) in
Eqs.(\ref{eq:E4.4}) we obtain the equations that eventually must
be solved. In these equations, according to (\ref{eq:E4.6}), there
is an operator $(\partial_{[3]} -{3\over 2}\partial_{[3]}\ln{\cal
R})f={\cal R}^{3/2}\partial_{[3]}({\cal R}^{-3/2}f)={\cal
R}\cdot{\cal R}^{3/2}\partial_{\rho}({\cal R}^{-3/2}f)$. Since
$\partial_A {\cal R}=\partial_A {\cal S}=\partial_A{\cal P}=
\partial_A{\cal Y} =0$ for $~A=0,1,2$, a simple
calculation with $\partial_\tau{\cal R}=\partial_\tau{\cal Y} =0$
yields the system,
\begin{eqnarray}\label{eq:E5.3}
{{\cal R}\over 2}
(\partial_{\rho}+\partial_{\tau})\big({u_R^2\over{\cal R}^3}\big)
+i{\cal R}\big({u_R^2\over{\cal R}^3}\big) (\partial_{\rho}+
\partial_{\tau}) \phi_R^u
= -i \big({u_L^2 \over{\cal R}^3}\big) e^{+i{\cal Y}_u},~~~{\rm (a)}
\nonumber\\* -{{\cal R}\over 2}
(\partial_{\rho}-\partial_{\tau})\big({u_L^2\over{\cal R}^3}\big)
-i{\cal R}\big({u_L^2\over{\cal R}^3}\big) (\partial_{\rho}-
\partial_{\tau})\phi_L^u=-i\big({u_R^2\over{\cal
R}^3}\big) e^{-i{\cal Y}_u},~~~{\rm (b)} \\*
e^{-i(\phi_L^u+\phi_R^u)}\bigg[\partial_{[1]} +i\partial_{[2]}\bigg]
u_R e^{i\phi_R^u}=0,~~~~~~~~~~~~~~~~~~~~~~~~~~~~~~~{\rm
(c)}\nonumber
\\* e^{-i(\phi_L^u+\phi_R^u)}\bigg[\partial_{[1]} +i\partial_{[2]}\bigg]u_L
e^{i\phi_L^u}=0,~~~~~~~~~~~~~~~~~~~~~~~~~~~~~{\rm (d)} \nonumber
\end{eqnarray}
where ${\cal Y}_u= \phi_L^u-\phi_R^u$. For the axial current directed inward,
in order to preserve an  intuitive physical understanding of a distance {\it from} an
object, we want ${\bm e}_{[3]}$ be directed outward. Then the
triplet $({\bm e}_{[1]}, {\bm e}_{[2]}, {\bm e}_{[3]})$ will be
left-handed. We have to take ${\cal J}^\mu=-{\cal R}e_{[3]}^\mu$,
$\Theta^\mu={\cal R} e_{[2]}^\mu$, and $\Phi^\mu={\cal R}
e_{[1]}^\mu$ in order for the vector product $[{\vec
e}_{[1]}\times {\vec e}_{[2]}]={\vec e}_{[3]}$ to represent
the external normal and the triplet $({\bm e}_{[1]}, {\bm
e}_{[2]}, {\bm e}_{[3]})$ to be right-handed. This results in the
interchange of the tetrad indices $1\leftrightarrow 2$ in
equations (\ref{eq:E2.3}), or, equivalently, in the change of the
sign of the tetrad components of the vector potential, $e{\tilde
A}_B\to -e{\tilde A}_B$. Thus, the string of the change of
variables becomes
\begin{eqnarray}\label{eq:E5.4}
\partial_{[0]}\to m{\cal R}\partial_{\tau}, ~\partial_{[3]}\to
-m{\cal R}\partial_{\rho},~eA_{[0]}\to{e\over m}{\cal R}A_\tau,~
eA_{[3]}\to-{e\over m}{\cal R}A_\rho,\nonumber\\~ e{\tilde
A}_{[0]} \to-{e\over m}{\cal R}{\tilde A}_\tau,~ e{\tilde
A}_{[3]}\to +{e\over m}{\cal R} {\tilde A}_\rho,~e{\tilde
A}_{[1]}\to -{e\over m}{\tilde A}_{[1]},~e{\tilde A}_{[2]}\to
-{e\over m}{\tilde A}_{[2]}.
\end{eqnarray}
Note, that in the course of the change of variables outlined above,
the sign of the $e{\tilde A}_{[3]}$ has been changed twice. Now,
using the physical variables (\ref{eq:E5.4}) in
Eqs.(\ref{eq:E3.5}) we arrive at a similar system,
\begin{eqnarray}\label{eq:E5.5}
-{{\cal R}\over 2}
(\partial_{\rho}-\partial_{\tau})\big({d_R^2\over{\cal R}^3}\big)
+i{\cal R}\big({d_R^2\over{\cal R}^3}\big) \bigg[(\partial_{\tau}-
\partial_{\rho})\phi_R^d-{2e\over m}(A_\tau-A_\rho)\bigg]
=-i\big({d_L^2\over{\cal R}^3}\big) e^{i{\cal Y}_d}, ~~~~~{\rm
(a)}\nonumber\\ {{\cal R}\over
2}(\partial_{\rho}+\partial_{\tau})\big({d_L^2\over{\cal R}^3}\big)
+i{\cal R}\big({d_L^2\over{\cal R}^3}\big) \bigg[(\partial_{\tau}+
\partial_{\rho}) \phi_L^d -{2e\over m}(A_\tau+A_\rho)\bigg]
= -i \big({d_R^2 \over{\cal R}^3}\big)
e^{-i{\cal Y}_d}, ~~~{\rm (b)}\nonumber\\
e^{-i(\phi_L^d+\phi_R^d)}\bigg[\partial_{[1]}+i\partial_{[2]}-
{2ie\over m}(A_{[1]}+iA_{[2]})\bigg] d_R e^{i\phi_R^d}=0,
~~~~~~~~~~~~~~~~~~~~~~~~~~~~~~{\rm (c)}\\~~~
~~e^{-i(\phi_L^d+\phi_R^d)}\bigg[\partial_{[1]} +i\partial_{[2]}-
{2ie\over m} (A_{[1]}+iA_{[2]})\bigg] d_L e^{i\phi_L^d}=0,
~~~~~~~~~~~~~~~~~~~~~~~~~~~~~~{\rm (d)}\nonumber
\end{eqnarray}
where ${\cal Y}_d= \phi_L^d-\phi_R^d$. The difference between
$\psi_u$ and $\psi_d$ is seen right in the equations of motion. The
tetrad components of an external field along holonomic coordinates,
$A_\tau, A_\rho \in S_{(03)}$, affect the $\psi_u$-mode not the
$\psi_d$-mode. Conversely, the associated with the non-holonomic
coordinates angular components $A_{[1]},A_{[2]} \in S_{(12)}$ are
assembled as the ladder operators and affect only $\psi_d$ pushing
it up to the state $\psi_u$. This difference between the last two
equations of systems (\ref{eq:E4.3}) and (\ref{eq:E5.5}) points to a
generic instability of the $\psi_d$- mode \footnote{Since ${\bm
e}_1$ and ${\bm e}_2$ are the  \textquotedblleft
angular\textquotedblright directions, it is instructive to recall
that the operators $L_+=L_1+ iL_2$ are ladder operators for the
angular momentum that moves eigenstate of the $L_z$ up. Both systems
(\ref{eq:E5.3}) and (\ref{eq:E5.5}) contain only $L_+$. While
$\psi_u$ cannot be pushed further up (and is stable), the $\psi_d$
is readily pushed up to the $\psi_u$. One can view these transitions
as a manifestation of the $\psi_d$-waveform's \textquotedblleft
motion\textquotedblright. In fact, it is a flow of surrounding Dirac
matter with ${\cal R}\geq 1$ that looks like a motion of the
$\psi_d$-dip (or void). }. It is discussed in   Sec.\ref{sec:Sec7} .

\subsectionn{Reduction to the real-valued functions}

As the last step before solving systems (\ref{eq:E5.3}) and
(\ref{eq:E5.5}) we split real and imaginary parts of the first two
equations of these systems and reduce equations to a form
convenient for finding the solutions. For the mode $\psi_u$ the
result reads as
\begin{eqnarray}\label{eq:E5.6}
{{\cal R}\over 2} \bigg({\partial\over\partial{\rho}}+
{\partial\over \partial{\tau}}\bigg)\big({u_R^2\over{\cal
R}^3}\big)= \big({u_L^2\over{\cal R}^3}\big) \sin{\cal
Y}_u,~~~~~~~~~~~~{\rm (a)}\nonumber\\{\cal R} \bigg({\partial\over
\partial{\rho}} +{\partial\over \partial{\tau}}\bigg) \phi_R^u= -
{u_L^2\over u_R^2} \cos{\cal Y}_u ,~~~~~~~~~~~~~~~ {\rm (a')}
\nonumber\\ {{\cal R}\over 2} \bigg({\partial\over\partial{\rho}}-
{\partial\over
\partial{\tau}}\bigg) \big({u_L^2\over{\cal R}^3}\big)=
\big({u_R^2\over{\cal R}^3}\big) \sin{\cal Y}_u,~~~~~~~~~~~{\rm (b)}
\nonumber\\ {\cal R}\bigg({\partial\over\partial{\rho}}-
{\partial\over \partial{\tau}}\bigg) \phi_L^u = {u_R^2\over u_L^2 }
\cos{\cal Y}_u .~~~~~~~~~~~~~~~~~{\rm (b')}
\end{eqnarray}
For the mode $\psi_d$ the result is somewhat different,
\begin{eqnarray}\label{eq:E5.7}
{{\cal R}\over 2} \bigg({\partial\over\partial{\rho}}-
{\partial\over \partial{\tau}}\bigg)\big({d_R^2\over{\cal
R}^3}\big)= -\big({d_L^2\over{\cal R}^3}\big) \sin{\cal Y}_d
,~~~~~~~~~~~~~~~~~~~~~~~~~~~~~~~~~~~{\rm (a)} \nonumber\\ {\cal R}
\bigg({\partial\over
\partial\rho} -{\partial\over \partial\tau}\bigg)\phi_R^d =
{d_L^2\over d_R^2}\cos{\cal Y}_d+{2e\over m}{\cal R}(A_\rho-A_\tau),
~~~~~~~~~~~~~~~~~~~~~~~~ {\rm (a')} \nonumber\\ {{\cal R} \over 2}
\bigg({\partial\over\partial{\rho}}+
{\partial\over\partial{\tau}}\bigg) \big({d_L^2\over{\cal
R}^3}\big)= -\big({d_R^2\over{\cal R}^3}\big) \sin{\cal Y}_d
,~~~~~~~~~~~~~~~~~~~~~~~~~~~~~~~~~~~{\rm (b)}\nonumber\\{\cal
R}\bigg({\partial\over
\partial\rho}+{\partial\over \partial\tau}\bigg)\phi_L^d
=-{d_R^2\over d_L^2 } \cos{\cal Y}_d+{2e\over m}{\cal R}
(A_\rho+A_\tau). ~~~~~~~~~~~~~~~~~~~~{\rm
(b')}
\end{eqnarray}
The phases $\phi_R^u$ and $\phi_L^u$ are affected in $\psi_u$ by the
right and left lightlike components of the vector potential,
respectively, but with the coupling constant $2e$. Conversely, the
phases $\phi_L^d$ and $\phi_R^d$ of the $\psi_d$ are not affected at
all.

Next, adding and subtracting equations (\ref{eq:E5.6}.$a'$) and
(\ref{eq:E5.6}.$b'$) and recalling that $\phi_L^u-\phi_R^u={\cal
Y}_u$ we find that
\begin{eqnarray}\label{eq:E5.8}
{\cal R}{\partial{\cal Y}_u\over\partial\rho}={\cal R}
{\partial{\cal Z}_u \over\partial\tau} +\bigg({\cal
X}_u^2+{1\over{\cal X}_u^2}\bigg) \cos{\cal Y}_u ,~~~~~~~~~~~~~~~{\rm (a)}
\nonumber\\ {d{\cal R}\over d\rho}=-\sin{\cal Y}_u,
~~~~~~~~~~~~~~~~~~~~~~~~~~~~~~~~~~~~~~~~~~~~~{\rm (b)}\nonumber\\ {\cal R}
{\partial{\cal Z}_u \over\partial\rho} =-\bigg({\cal
X}_u^2-{1\over{\cal X}_u^2}\bigg)\cos{\cal Y}_u.~~~~~~~~~~~~~~~~~~~~~~~~~{\rm
(c)}\\ \big[\partial_{[1]}+i\partial_{[2]}\big]{\cal Z}_u
=0,~~~~~~~~~~~~~~~~~~~~~~~~~~~~~~~~~~~~~~~{\rm (d)} \nonumber
\end{eqnarray}
where ${\cal Z}_u=\phi_L^u +\phi_R^u$ and $u_L/u_R={\cal X}_u $.
Repeating the same for the mode $\psi_d$ we obtain,
\begin{eqnarray}\label{eq:E5.9}
-{\cal R}{\partial{\cal Y}_d\over
\partial\rho}={\cal R} {\partial{\cal Z}_d \over\partial\tau}
+\bigg({\cal X}_d^2+{1\over{\cal X}_d^2}\bigg)\cos{\cal Y}_d
-{4e\over m} {\cal R} A_\tau,~~~ {\rm (a)}\nonumber\\ {d{\cal R}
\over d\rho}=+\sin{\cal Y}_d,
~~~~~~~~~~~~~~~~~~~~~~~~~~~~~~~~~~~~~~~~~~~~{\rm (b)} \nonumber\\
{\cal R}{\partial{\cal Z}_d\over \partial\rho}= -\bigg({\cal X}_d^2
-{1\over{\cal X}_d^2}\bigg)\cos{\cal Y}_d +{4e\over m}{\cal
R}A_\rho.~~~~~~~~~~~~~{\rm (c)}\\
\big[\partial_{[1]}+i\partial_{[2]}\big]{\cal Z}_d ={4e\over m}
(A_{[1]}+iA_{[2]}),~~~~~~~~~~~~~~~~~~~{\rm (d)}\nonumber
\end{eqnarray}
where ${\cal Z}_d=\phi_L^d +\phi_R^d$  and $d_L/d_R={\cal X}_d $.
Eqs.(\ref{eq:E5.8}.d) and (\ref{eq:E5.9}.d) are easily obtained from
Eqs.(\ref{eq:E5.3}.c,d) and (\ref{eq:E5.5}.c,d) because none of the
amplitudes
$u_R,u_L$ and $d_R,d_L$ and  of the phase differences
${\cal Y}_u,{\cal Y}_d$ depend on the angular variables $S^{[1]}$ and $S^{[2]}$.
We postpone discussion of the Eqs.(\ref{eq:E5.3}.c,d) and
(\ref{eq:E5.5}.c,d), which are responsible for the stability or
instability of the solutions, till Sec.\ref{sec:Sec7}.

Before looking for the stationary modes of the nonlinear Dirac
equation we are going to learn whether they can emerge as asymptotic
configurations at $\tau\to\infty$ of a transient process that can
begin from an arbitrary perturbation or are they { \it ad hoc}
constructed isolated solutions. By adding and subtracting
Eqs.(\ref{eq:E5.6}.a,b), with the l.h.s. reduced to the logarithmic
derivatives, and some simple algebra we obtain
\begin{eqnarray}\label{eq:E5.10}
{\partial{\cal X}_u\over\partial\tau}={\cal X}_u\bigg({\cal
X}_u-{1\over{\cal X}_u}\bigg)^2~{\partial\ln{\cal
R}\over\partial\rho},~~~~{\partial{\cal X}_u\over\partial\rho}={\cal
X}_u\bigg({\cal X}_u^2-{1\over{\cal X}_u^2}\bigg)~{\partial\ln{\cal
R}\over\partial\rho}, 
\end{eqnarray}
where $\partial_\rho\ln{\cal R}=-\sin{\cal Y}_u/{\cal R}$. Excluding
from these two equations the $\partial_\rho\ln{\cal R}$, one finds a
first-order wave equation, $\partial_\tau {\cal X}+c({\cal
X})\partial_\rho {\cal X}=0$, with the wave velocity $c({\cal
X})=(1- {\cal X}^2)/(1+ {\cal X}^2)$. Because $c(1)=0$, the
\textquotedblleft propagation\textquotedblright of ${\cal X}$  stops
at ${\cal X}=1$. Since ${\cal R}$ depends only on $\rho$, both
equations (\ref{eq:E5.10}) are easily integrated,
\begin{eqnarray}\label{eq:E5.11}
{\cal X}_u^2(\tau,\rho)=1-{1\over 2\partial_\rho\ln{\cal
R}\cdot\tau+C_2(\rho)},~~~~ {\cal X}_u^2(\tau,\rho)={{\cal
R}^4-C_1(\tau)\over {\cal R}^4+C_1(\tau)}, 
\end{eqnarray}
where the constants of integration $C_1(\tau)$ and $C_2(\rho)$ are
arbitrary functions of only one argument. Since ${\cal
X}_u(\infty,\rho)=1$ (and then $C_1(\infty)=0$), we find that at the
asymptotic world time $\tau$ the coefficients in front of $\cos{\cal
Y}_u(\rho)$ in Eqs.(\ref{eq:E5.8}.a) and  (\ref{eq:E5.8}.c) become
$2$ and $0$, respectively. Assuming further that $e=0$ (no external
field), we find that $\partial_\rho{\cal Z}_u=0$ and thus  ${\cal
Z}_u={\cal Z}_u(\tau)$. Now, $\partial_\tau{\cal Z}_u$ is the only
potentially $\tau$-dependent term in Eq.(\ref{eq:E5.8}.a); then it
cannot depend on $\tau$. Therefore, the only option is
$\partial_\tau{\cal Z}_u= -2E=const$,    ${\cal Z}_u=-2E\tau$, and
it immediately follows that $u_L^2=u_R^2=u^2={\cal R}/2$ (which is
an evidence that the particle is at rest!). Eqs.(\ref{eq:E5.11}) are
compatible only in the limit of $\tau\to\infty$ since they imply
$\partial_\tau{\cal R}=0$; a transient process naturally requires
that  $\partial_\tau{\cal R}\neq 0$. Similar results are true for
the mode $\psi_d$.

\sectionn{Stationary solutions.}\label{sec:Sec6}
\renewcommand{\theequation}{6.\arabic{equation}}
\setcounter{equation}{0}
{ \fontfamily{times}\selectfont
\noindent

Being interested here only in stationary states we assume a
trivial dependence of the phases of Dirac field components on
$\tau$, $\psi\propto e^{-iE\tau}$,  and replace,
$\phi_R\to\phi_R(\rho)-E\tau$, $\phi_L\to\phi_L(\rho)-E\tau$,
throughout this section. Then,  $u_L^2=u_R^2=u^2={\cal R}/2$
and $d_L^2=d_R^2=d^2={\cal R}/2$. Taking further the coupling constant
$e=0$, which is, in fact, equivalent to a one-body approximation,
we end up with an autonomous system of two ODEs for two unknown
functions (the amplitude ${\cal R}(\rho)$ and the phase difference
${\cal Y}(\rho)$) of the natural parameter $\rho$ (and not the
affine parameter $s$!) along the radial geodesic lines.
\subsectionn{Localized solution for the ${\bm \psi_u}$-mode of the Dirac field.}

In the stationary case, Eqs.(\ref{eq:E5.8}) for the $\psi_u$-mode
with the axial current directed outward, read as
\begin{eqnarray}\label{eq:E6.1}
{\cal R}(\rho){d{\cal Y}_u(\rho)\over d\rho}=-2\epsilon{\cal
R}(\rho) +2\cos{\cal Y}_u(\rho),~~~{\rm (a)}~~~~ {\cal
R}(\rho){d{\cal R}(\rho) \over d\rho}=-{\cal R}(\rho)\sin{\cal
Y}_u(\rho)~~~{\rm (b)}~,
\end{eqnarray}
where $\epsilon=E/m$. The characteristic equation for this system,
\begin{eqnarray}\label{eq:E6.2}
{d{\cal Y}_u\over -2\epsilon{\cal R}+2\cos{\cal Y}_u}=-{d{\cal R}
\over {\cal R}\sin{\cal Y}_u},
\end{eqnarray}
is easily solved in terms of $w({\cal R}) =\cos{\cal Y}_u$. Then,
${\cal R}w'_{\cal R} -2w +2\epsilon{\cal R}=0$, and
\begin{eqnarray}\label{eq:E6.3}
\cos{\cal Y}_u = C{\cal R}^2+2\epsilon {\cal R},~~~
\end{eqnarray}
is the first integral of system (\ref{eq:E6.1}) depending on
one, yet undetermined, constant $C$.

~{\bf 1. General (periodic) solution.} Solving Eq.(\ref{eq:E6.3})
for ${\cal R}$, and taking into account two possible signs of $C$,
one can rewrite Eq.(\ref{eq:E6.1}a) as
\begin{eqnarray}\label{eq:E6.4}
{d{\cal Y}\over d\rho}=\mp 2\sqrt{C}\cdot\sqrt{{\epsilon^2\over C}
+\cos{\cal Y}},~~~C>0~~{\rm and}~~{d{\cal Y}\over d\rho}=\mp
2\sqrt{|C|}\cdot\sqrt{{\epsilon^2\over |C|} -\cos{\cal Y}}.~~~C<0.
\end{eqnarray}
Thus, the dependence $\rho({\cal Y})$ in the cases $C>0$ and $C<0$
is given by the quadratures \cite{Elliptic},
\begin{eqnarray}\label{eq:E6.5}
\rho({\cal Y}) = {\mp 1\over \sqrt{C(1+b^2)}}\int_{0}^{{\cal Y}/2}
{~d\phi \over \sqrt{1- {2\over 1+b^2} \sin^2{\phi}}} = {\mp 1\over
\sqrt{C(1+b^2)}}F\big({{\cal Y}\over 2}\big|{2\over 1+b^2}~\big)],
~C>0 ,~~~\\\label{eq:E6.6} \rho({\cal Y})= {\mp 1\over
\sqrt{|C|(b^2-1)}} \int_{0}^{{\cal Y}/2} {~d\phi \over \sqrt{1-
{2\over 1-b^2}\sin^2{\phi}}} = {\mp 1\over
\sqrt{C(b^2-1)}}F\big({{\cal Y}\over 2}\big|{2\over 1-b^2}~\big)0,
~C<0 ,~~~
\end{eqnarray}
where $b^2=\epsilon^2/|C|>0$ and $w=F(\Phi|k^2)={\rm sn}^{-1}
(\sin\Phi|k^2) $ is the incomplete elliptic integral of the first
kind\footnote{These expressions have no practical value and will
be used below for a sole purpose of proving that the modules of the
elliptic integrals must equal +1 by the physics of the problem.
Then, and only then is ${\cal R}(\rho)$ not oscillating in radial
direction. This uniquely fixes the constant as $|C|=\epsilon^2$
and guarantee that elliptic integrals become smooth elementary
functions. The limits of integration in (\ref{eq:E6.5}) are
tentative.},
\begin{eqnarray}\label{eq:E6.7}
F(\Phi|k^2)=\int_0^\Phi(1-k^2 \sin^2\phi)^{-1/2}d\phi
=\int_0^X[(1-x^2)(1-k^2x^2)]^{-1/2}dx,~~~X=\sin\Phi.
\end{eqnarray}
Its inverse is a well-known Jacobi's amplitude function, $\Phi={\rm
am}(w|k^2)$. Leaving aside for a while the case of $C<0$, we
readily find that
\begin{eqnarray}\label{eq:E6.8}
\sin{{\cal Y}\over 2} = {\rm sn}(u|k^2),~~\cos{{\cal Y}\over 2}=
{\rm cn} (u|k^2), ~~\sin{\cal Y}=2{\rm sn}(u|k^2){\rm
cn}(u|k^2),~~\cos{\cal Y}={\rm cn}^2(u|k^2)-{\rm sn}^2(u|k^2),
\end{eqnarray}
where $u=\sqrt{\epsilon^2+C}~\rho =F({\cal Y}/2|2/(1+b^2))$,
$k^2=2/(1+b^2)$. Now Eq.(\ref{eq:E6.1}b) becomes,
\begin{eqnarray}\label{eq:E6.9}
{d{\cal R}(\rho) \over d\rho}=-\sin{\cal Y}(\rho)=-2{\rm
sn}(u|k^2) {\rm cn}(u|k^2),
\end{eqnarray}
and, since $\int{\rm sn}(u|k^2){\rm cn}(u|k^2)du=-{\rm
dn}(u|k^2)/k^2$~\cite{Elliptic}, the latter equation is readily
integrated,
\begin{eqnarray}\label{eq:E6.10}
{\cal R}(\rho) ={\sqrt{\epsilon^2+C}\over C}~{\rm dn}
\big(\sqrt{\epsilon^2+C}~\rho \big|{2\over 1+b^2}\big),~~ C>0.
\end{eqnarray}
In the second case of $C<0$ we would have
\begin{eqnarray}\label{eq:E6.11}
{\cal R}(\rho) ={\sqrt{\epsilon^2-|C|}\over |C|}~{\rm dn}
\big(\sqrt{\epsilon^2-|C|}~\rho \big|{2\over 1-b^2}\big) ,~~ C<0.
\end{eqnarray}
The Jacobi's elliptic functions ${\rm sn}(u|k^2)$, ${\rm
cn}(u|k^2)$ and ${\rm dn}(u|k^2)$ are known to be double-periodic
functions of their argument. While periodic behavior of the phase
${\cal Y}(\rho)$ cannot {\it a priori} be excluded, {\it
periodicity in radial direction is impossible for the invariant
density ${\cal R}(\rho)$}, simply because it would conflict with the
physical localization.

~{\bf 2. Localized (aperiodic) solution.} There is, however, a
special case when the module of the elliptic function $k=1$ and
the periodicity disappears (the period becomes infinite). For the
Eq.(\ref{eq:E6.10}), this means that $b^2=\epsilon^2/|C|=1$ so
that ${\rm dn}(u|1)=1/\cosh{u}$ (as well as ${\rm
cn}(u|1)=1/\cosh{u}$ and ${\rm sn}(u|1)=\tanh{u}$). For the
Eq.(\ref{eq:E6.11}) the same would mean $b^2=-1$, which is
impossible, since $b^2>0$, by definition. Hence, the case of $C<0$
must be dropped from further consideration.

The constant $C$ of integration in the Eq.(\ref{eq:E6.3}) is now
uniquely determined as $C=\epsilon^2= (E/m)^2$, and the equation of
characteristics of system (\ref{eq:E6.1}) becomes
\begin{eqnarray}\label{eq:E6.12}
\cos{\cal Y}+1 =2\cos^2({\cal Y}/2)= (\epsilon{\cal R}+1)^2.
\end{eqnarray}
Since the Jacobi's elliptic functions with module $k=1$ are
elementary functions, it is much easier to return to the original
system (\ref{eq:E6.1}) and the characteristic equation
(\ref{eq:E6.12}) with $C=\epsilon^2$, using the latter as a
constraint. After using the constraint (with the signs to be
determined later), $\epsilon {\cal R}+1= \pm\sqrt{2} \cos({\cal
Y}_u/2)$, the system (\ref{eq:E6.1}) simplifies to
\begin{eqnarray}\label{eq:E6.13}
{d{\cal Y}_u\over d\rho}= -2^{3/2}\epsilon\cos{{\cal Y}_u\over
2},~~~{\rm (a)}~~~~ {d{\cal R} \over d\rho}=-\sin{\cal Y}_u=
-2\sin{{\cal Y}_u\over 2}\cos{{\cal Y}_u\over 2}, ~~~{\rm (b)}~
\end{eqnarray}
and its first equation is readily integrated to $\rho({\cal Y})$
first, and then yields ${\cal Y}(\rho)$
\begin{eqnarray}\label{eq:E6.14}
\sqrt{2}\epsilon\rho=\tanh^{-1}(\sin {{\cal Y}_u\over
2}),~~\sin{{\cal Y}_u\over 2} =-\tanh(\sqrt{2}
~\epsilon\rho),~~\cos{{\cal Y}_u\over 2}= {1\over\cosh(\sqrt{2}
~\epsilon\rho)}.
\end{eqnarray}
When $\rho\to\infty$, we have $\epsilon {\cal R}+1\to 0$, which is
possible only when $\epsilon =E/m <0$. We also obtain the
anticipated $\sin{\cal Y}(\infty) = 0$ and $\cos{\cal
Y}_u(\infty)= -1$, i.e. ${\cal Y}_u(\infty)=\pi$. Returning the
result of integration into Eqs.(\ref{eq:E6.12}) and
(\ref{eq:E6.13}b), we simplify the latter to
\begin{eqnarray}\label{eq:E6.15}
\epsilon {\cal R}+1= {-\sqrt{2} \over\cosh(\sqrt{2}
~\epsilon\rho)},~~~ {d{\cal R}\over d\rho}=-\sin{\cal Y}_u(\rho) =
-2{\sinh(\sqrt{2}|\epsilon|\rho) \over
\cosh^2(\sqrt{2}\epsilon\rho)}.
\end{eqnarray}
In order for this solution to be interpreted as an isolated
particle at rest, we must require that $E=-m$. Thus the solution
\begin{eqnarray}\label{eq:E6.16}
{\cal R}(\rho)={\sqrt{2}\over \cosh(\sqrt{2}\rho)}+1,
\end{eqnarray}
is the mode with the negative energy  with respect to the vacuum
level zero attributed to ${\cal R}=1$. Finally, in natural units,
\begin{eqnarray}\label{eq:E6.17}
\sin{{\cal Y}_u\over 2} =\tanh(\sqrt{2} ~m\rho),~~ {\cal R}(\rho)=
{\sqrt{2}\over \cosh(\sqrt{2}m\rho)}+1.
\end{eqnarray}
This result also follows from Eq.(\ref{eq:E6.9}), since ${\rm
dn}(u|1)=1/\cosh u$. We can take the radius $\rho_0$ of the
spherical surface, where $d{\cal R}/d\rho$ reaches its maximum
(the inflection point) for the size of the particle. Here,
$\sin{\cal Y}_u(\rho_0) =1$, and, consequently, $\sinh(\sqrt{2}
~m\rho_0)=1$, $\cosh(\sqrt{2} ~m\rho_0)=\sqrt{2}$. Therefore (in
natural units), $$\rho_0={\sinh^{-1}(1)\over
\sqrt{2}~m}={0.623\over m}~~{\rm and}~~ s_0= {\rho_0 \over {\cal
R}(\rho_0)} ={1\over m},$$ as it was previously contemplated.  At
the radius $\rho_0$, also as expected, the phase is  ${\cal
Y}_u(\rho_0)=\pi/2$.  Indeed, $\cos({\cal Y}_u(\rho_0)/2)=
1/\sqrt{2}=\cos(\pi/4)$ and $\sin{\cal Y}_u(\rho_0)=1$, ${\cal
R}_u(\rho_0)=2$. The peak amplitude ${\cal R}_u(0)=1+\sqrt{2}$.

\subsectionn{Dirac field in $\psi_d$-mode.}

We expect that in real world the mode $\psi_d$ with the axial
current looking inward will be unstable and not similar, even
qualitatively, to the mode $\psi_u$. However, it is instructive to
repeat the previous steps and consider only Eqs.(\ref{eq:E5.7})
leaving aside Eqs.(\ref{eq:E5.5}.c,d). Then most of the analysis
remains the same and only Eqs.(\ref{eq:E6.1}) and
(\ref{eq:E6.12})-(\ref{eq:E6.16}) are modified.
Eqs.(\ref{eq:E6.1}) now read as
\begin{eqnarray}\label{eq:E6.18}
{\cal R}(\rho){d{\cal Y}_d(\rho)\over d\rho}=+2\epsilon{\cal
R}(\rho) -2\cos{\cal Y}_d(\rho),~~~{\rm (a)}~~~~{\cal R}(\rho)
{d{\cal R}(\rho) \over d\rho}=+{\cal R}(\rho)\sin{\cal
Y}(\rho)~,~~~{\rm (b)}
\end{eqnarray}
and the change of the sign of $\epsilon$ and of the slope does not
affect the characteristic equation (\ref{eq:E6.3}) except that
we must replace $\epsilon\to -\epsilon,\cos{\cal Y}_u\to
-\cos{\cal Y}_d $ in it. Then the cases $C>0$ and $C<0$ must be
swapped in Eqs.(\ref{eq:E6.4})-(\ref{eq:E6.11}) with the
conclusion that  constant $C$ must be determined as
$C=-\epsilon^2= -(E/m)^2$, and equation (\ref{eq:E6.3}) of
characteristics of system (\ref{eq:E6.18}) reads as
\begin{eqnarray}\label{eq:E6.19}
1-\cos{\cal Y} =2\sin^2({\cal Y}/2)= (1-\epsilon{\cal R})^2.
\end{eqnarray}
After using the constraint, $1-\epsilon {\cal R}= -\sqrt{2}
\sin({\cal Y}_d/2)$, the system (\ref{eq:E6.18}) becomes,
\begin{eqnarray}\label{eq:E6.20}
{d{\cal Y}_d\over d\rho}= -2^{3/2}|\epsilon|\sin{{\cal Y}_d\over
2},~~~{\rm (a)}~~~~ {d{\cal R} \over d\rho}=\sin{\cal Y}_d=
2\sin{{\cal Y}_d\over 2}\cos{{\cal Y}_d\over 2}, ~~~{\rm (b)}~
\end{eqnarray}
and its first equation is readily integrated as
\begin{eqnarray}\label{eq:E6.21}
\sqrt{2}m\rho=\tanh^{-1}\big(\cos {{\cal Y}_d\over
2}\big),~~\cos{{\cal Y}_d\over 2} =\tanh(\sqrt{2}
~\epsilon\rho),~~\sin{{\cal Y}_d\over 2}= {1\over\cosh(\sqrt{2}
~\epsilon\rho)}.
\end{eqnarray}
Acting as previously, we simplify the constrain and
Eq.(\ref{eq:E6.20}.b) to
\begin{eqnarray}\label{eq:E6.22}
1-\epsilon {\cal R}= {\sqrt{2} \over\cosh(\sqrt{2}
~\epsilon\rho)},~~~ \sin{\cal Y}_d(\rho) =
+2{\sinh(\sqrt{2}\epsilon\rho) \over
\cosh^2(\sqrt{2}\epsilon\rho)}={d{\cal R}\over d\rho},
\end{eqnarray}
where the second equation is identical to (\ref{eq:E6.18}.b) and
is a consequence of the first one. When $\rho\to\infty$, we have
$1-\epsilon {\cal R}\to 0$, which is possible only when $\epsilon
=E/m >0$. Here, the condition of a particle at rest requires  that
$\epsilon=E/m=+1$. We also obtain the anticipated $\sin{\cal
Y}(\infty) = 0$ and $\cos{\cal Y}_d(\infty)= 1$, i.e. ${\cal
Y}_d(\infty)=\pi$. Thus the solution (in natural units)
\begin{eqnarray}\label{eq:E6.23}
{\cal R}(\rho)=1-{\sqrt{2}\over \cosh(\sqrt{2}m\rho)},
\end{eqnarray}
can be interpreted as an isolated particle at rest with the
positive energy $E=+m$, which is $2m$ higher than that for the similar
localized static $\psi_u$-mode. Here, once again, ${\cal R}
(\infty)=1$. If the auto-localization is a real process it must
favor localization  not of $\psi_d$ that has a dip , but the bump
of $\psi_u$. This is also a hint that an {\it ad hoc} created
$\psi_d$ can be unstable (as it is in Nature). We elaborate on it
in the last section.

Finally, for the mode with a dip of the invariant density in its
interior, the invariant density reaches its theoretical minimum,
${\cal R}(\rho_0)=0$, at the inflection point $\rho_0=1/m$ \footnote{In general, none of the Dirac currents vanishes at ${\cal R}=0$; they all become proportional to one lightlike vector that must have both up- and down-components. Then nothing can  identify the surface $S_{(12)}$ of constant $\tau$ and $\rho$ as a two-dimensional sphere.}. At this point we have $\sin{\cal Y}_d(\rho_0)=1$, i.e. ${\cal
Y}_d(\rho_0) =\pi/2$. Inside this radius the density ${\cal R}$,
as formally defined by (\ref{eq:E6.23}), becomes negative, which
is impossible.  This can be a yet another indication that {\it an
isolated localized } negative charge is unstable (at least in the
absence of external field or of stable third bodies nearby). In
other words, even being localized, it most likely is
\textquotedblleft an agile shallow deepening on a hill\textquotedblright.
Indeed, in real world of a stable matter, all electrons are light and only weakly
localized around atomic nuclei, so that normal matter is charge-neutral. The heavy inward-polarized
particles (e.g., antiprotons) are found only rarely and they would not be detected
without abundant normal matter nearby. These probably are
\textquotedblleft deep holes on a high hill\textquotedblright. Verification of this
hypothesis is not a one-body problem and is beyond the scope of
this work.

\sectionn{Stability and an effective Lagrangian}\label{sec:Sec7}
\renewcommand{\theequation}{7.\arabic{equation}}
\setcounter{equation}{0}
{ \fontfamily{times}\selectfont
\noindent

The two exact solutions of the Dirac equation in one-body
approximation, given by Eqs.(\ref{eq:E6.13}) -(\ref{eq:E6.16}) for
the modes $\psi_u$, and by Eqs.(\ref{eq:E6.21}) -(\ref{eq:E6.23})
for the mode $\psi_d$, seem to be very similar to each other
except that $\psi_u$ has a bump and $\psi_d$ has a dip of the
invariant density near the center. According to the initial
hypothesis, they correspond to positive and negative charges,
respectively. The primary guess was \cite{PartI,Localization} that the former
must be localized better and (if being unstable) live longer than
the later, solely because the proper time in their interior flows
the slower, the higher the invariant density is. Beyond the
one-body approximation, the difference between these solutions is
encoded mainly in the last two equations of the system
(\ref{eq:E5.3}) for $\psi_u$ and (\ref{eq:E5.5}) for $\psi_d$. In
the case of $\psi_u$ they do not depend on the external field
$A_\mu$, while in the case of $\psi_d$ they do. Furthermore, the
tetrad components $A_{[1]}+ iA_{[2]}$ in Eqs.(\ref{eq:E5.5}.c,d)
oscillate with time as $e^{-2im\tau}$ and can cause a transition
from $\psi_d$ to $\psi_u$.

The field $A_\mu$ in the Dirac equation is an external field.
Remarkably, whatever this field is, the Dirac field determines world time
across every auto-localized object. In a sense, all
solutions of Eqs.(\ref{eq:E5.3}) and (\ref{eq:E5.5}) with the
energy $\epsilon=E/m$ are the static solutions. But it is
well-known that not all static solutions are stable. Solutions
(\ref{eq:E6.16}) and (\ref{eq:E6.23}) obtained in absence of an
external field are both truly static since there is nothing in
Eqs.(\ref{eq:E6.1}) and (\ref{eq:E6.18}) that could have trigger
instability. To investigate the effects of instability one must
return to Eqs.(\ref{eq:E5.5}.c,d) and also to Eqs.(\ref{eq:E5.8})
and (\ref{eq:E5.9}), which also account for the external field
$A_D$ and dynamics of the sums of  phases, ${\cal Z}= \phi_L+
\phi_R$. The problem has two different aspects, {\it viz.},
formation of a perturbation and its decay.

Below, we try to specify both aspects and speculate regarding
possible approaches/tools. The following terminology seems most
appropriate for the discussion. Let us consider the components of
$\psi_u$ and $\psi_d$ as the wave functions of the initial state and
denote them as $|u\rangle_i,|d\rangle_i$. Next, let us contract
Dirac equation with the Hermit conjugated wave function of a
\textquotedblleft final state\textquotedblright, $_f\!\langle u|,
_f\!\langle d|$ and consider $_f\!\langle ... \rangle_i$ as
\textquotedblleft transition amplitudes\textquotedblright.

\subsectionn{ Creation of perturbations in Dirac vacuum.}

The problem of what may trigger the initial (and almost necessarily
unstable) configuration is the most subtle one. Classically, one has
to start with arbitrary initial field $\psi$ and a plausible
external field $A_\tau\pm A_\rho$ (i.g., of the cosmic microwave
background). In quasi-static regime, the interaction of  reasonably
well defined initial states  $|u\rangle_i$ with the lightlike
components $A_\tau\pm A_\rho$ of the vector potential is not
distracting, since Eqs.(\ref{eq:E5.3}.a,b) can contribute only to
diagonal (with respect to the spin) matrix elements,
\begin{eqnarray} \label{eq:E7.1}
-4ie _f\!\langle d_R|(A_\tau+ A_\rho )|d_R\rangle_i,~~~~ -4ie
_f\!\langle d_L|(A_\tau-A_\rho )|d_L\rangle_i.
\end{eqnarray}
These are {\it not} the transitions between  up- and down-states.
Regardless  how weak this interaction is, it takes place in enormous
space and for astronomical times. It {\it can} collapse to a
solitary excitation just because such excitations exist. This
mechanism can be considered as a potential source of the cosmic
positron excess (for an extensive review see Ref. \cite{Serpico}).
Furthermore, in Eqs.(\ref{eq:E5.3}.c,d) that could have trigger
transition from up- to down- states, there is no interaction terms
at all. Thus, solution (\ref{eq:E6.17}) of Eqs.(\ref{eq:E5.3}),
which is associated with a positive charge, is expected to be
stable.

\subsectionn{ Decay of an initial perturbation.}

If an initial finite waveform is given, a reasonable theory must
predict its decay into stable solitary configurations.
Eqs.(\ref{eq:E5.5}.c,d) (unlike (\ref{eq:E5.3}.c,d)) prompt the
interaction
\begin{eqnarray} \label{eq:E7.2}
-4ie _f\!\langle u_R|e^{i(\phi_L^d+\phi_R^d)}(A_{(1)}+iA_{(2)} )
|d_R\rangle_i~~~{\rm and}~~-4ie _f\!\langle
u_L|e^{i(\phi_L^d+\phi_R^d)}(A_{(1)}+iA_{(2)}) |d_L\rangle_i~,
\end{eqnarray}
that affects stability of the inward-polarized state and causes its
flip into a stable up-state. In these formulae, $A_{(1)}$ and
$A_{(2)}$ are the components of vector potential with respect to a
judiciously chosen basis $({\bm h}_1,{\bm h}_2)$ on the surface
$S_{(12)}\in \mathbb{M}$ mapped onto $\mathbb{R}^4$. The transition
from unstable mode to the stable one is due to the charged Dirac
currents that naturally oscillate as $e^{-2im\tau}$, and this
transition can be triggered by almost any external electromagnetic
field.  The latter can be random or regular and originate, e.g.,
from the cosmic background. Possibly, they can even stabilize the
$\psi_d$ mode for a long time. This could explain the difference
between an apparently stable particle in a storage ring and a
visibly unstable particle in the natural world.

\subsectionn{ Similarity to magnetic resonance?}

The matrix elements (\ref{eq:E7.2}) are intimately connected with
the dynamics of the spin $1/2$ in magnetic field, where quantum and
classical equations of motion coincide. Indeed, the sectional
curvature\footnote{The sectional curvature of a surface spanned by a
net of the lines of the vectors ${\bm e}_1$ and ${\bm e}_2$ equals
to the angle by which the basis $({\bm e}_1,{\bm e}_2$) is rotated
after moving along an infinitesimal loop within this surface.} of
the spherical surface $S_{(12)}$ (the curvature of the lines of the
charged currents ${\bm \Theta}$ and ${\bm \Phi}$),
\begin{eqnarray} \label{eq:E7.3}
R^t_{1212}=2e(\partial_{[1]} A_{[2]}-\partial_{[2]} A_{[1]})-
4e^2(A_{[1]}^2+ A_{[2]}^2) =2eF_{12}=2eB_{[3]},
\end{eqnarray}
is totally due to the projection of the {\it external magnetic
field} onto radial direction of the axial current. If such a
projection is not zero, it will cause flip of the spin
polarization into the outward direction of the stable
$\psi_u$-mode.

\subsectionn{  An effective Lagrangian.}More accurate approach that
would allow one to go beyond the lowest order approximation can
probably be based on the so-called effective Lagrangian,
$\mathcal{L}=\psi^+[i\alpha^A D_A\psi- m\rho_1 ] \psi$, with the
operator of Eq.(\ref{eq:E2.2}) in brackets. The terms depending on
$A_\mu$ in it can be viewed as the interaction with the outside
sources. Retaining the interaction term ($e\neq 0$), actually,
leads beyond the one-body approximation.  Below, solely for the
purpose of stability analysis, we add the alien up- and/or
down-components as a perturbation. The state is supposed to be
stable if the alien components dissipate due to the interaction.
It will be genuinely unstable if the interaction enforces
dissipation of the native components.  We continue to dub the
configurations with $u_L^2+u_R^2>d_L^2+d_R^2$ as $\psi_u$ (with
native $u$ and an admixture of alien $d$). Those with
$u_L^2+u_R^2<d_L^2+d_R^2$ are dubbed as $\psi_d$ (with native $d$
and alien $u$).

Let us look at the terms associated with the charged currents
$\Theta^a$ and $\Phi^a$ and consider the matrix element,
\begin{eqnarray} \label{eq:E7.4}
T_{ab}=<b| T |a>= \psi^+_b\bigg[\alpha^{[1]} \big( - ieA_{[1]}-
e\tilde{A}_{[2]}\big)\psi + \alpha^{[2]} \big(- ieA_{[2]}+
e\tilde{A}_{[1]} \big)\bigg]\psi_a,~~~
\end{eqnarray}
between the configurations $\psi_a$ and $\psi_b$. Here,
$\tilde{A}_D$ stands for  $A_D$ when the triplet $({\bm e}_{[1]},
{\bm e}_{[2]}, {\bm e}_{[3]})$ forms the right-handed system, and
for $-A_D$ when this triplet is left-handed. As an illustration,
consider a particular term assuming native $u_L,u_R$ and alien
$d_L,d_R$; then $T_{ab}$ is
\begin{eqnarray}\label{eq:E7.5}
\psi_b^+T_+\psi_a=\psi_b^+[eA_{[1]}(-i\alpha^{[1]}+\alpha^{[2]})+
eA_{[2]}(-i\alpha^{[2]}-\alpha^{[1]})]\psi_a\\
=-2ie(A_{[1]}-iA_{[2]}) \cdot \psi_b^+\rho_3
\sigma^+\psi_a~~~~~~~~~~~~~~~~~~~~~~~\nonumber\\= -2e A_\mu
(e_{[1]}^\mu -ie_{[2]}^\mu)\psi_b^+ {\cal O}_u\rho_3 \sigma^+{\cal
O}_d\psi_a.~~~~~~~~~~~~~~~\nonumber
\end{eqnarray}
Here, $\sigma^+=(\sigma^{[1]}+ i\sigma^{[2]})/2$ is the ladder
(spin-flip) operator for the projection of spin 1/2 onto the
positive direction ${\bm e}_{[3]}$ of the right-hand oriented
triplet.  Let us recall that ${\cal O}_{u/d}= (1\pm\sigma^3)/2$
are the projection operators onto the up-/down-components of the
Dirac spinor. In detail, the action of the operator $T_+$ is as
follows. The ladder operator $\rho_3\sigma^+$ eliminates the
native components $u_R$ and $u_L$ (acting on $\psi_a$ as ${\cal
O}_d$) and replaces them with the alien $d_R$ and $d_L$, producing
$\psi'=(d_R,0,-d_L,0)$. Since $\sigma^+{\cal O}_d= \sigma^+$, this
can be viewed as a two-step action. Namely, the ${\cal O}_d$
(inherited from connection (\ref{eq:E2.4})) filters out the
$d_R$ and $d_L$ in their alien position, and then $\sigma^+$ moves
them \textquotedblleft up\textquotedblright, thus filtering out the positive
helicity of the native \textquotedblleft up\textquotedblright-final state
$\psi_b^+{\cal O}_u$. In other words, $\psi_b^+T_+\psi_a\propto
(u^*_{bR} d_{aR}-u^*_{bL} d_{aL}) e^{-2im\tau}$. If the state
$\psi_a$ was a pure up-state $\psi_u$ and had no components
$(d_R,d_L)$ at all, then $\psi_b^+T_+\psi_u=0$; this is the case
of Eqs.(\ref{eq:E4.3})-- the $\psi_u$ does not interact with the
external $A_{[1]},A_{[2]}$. Conversely, the state $\psi_d$ that
has only $(d_R,d_L)$ is unstable under this interaction and the
charged currents will convert it to the up-state. This reproduces
the primitive analysis of Eqs.(\ref{eq:E7.1}) and (\ref{eq:E7.2}).


Since the effective Lagrangian is nonlinear, there are many open
questions, which cannot be addressed comprehensively within the
scope of the present work. For example, it is not clear {\it a
priori}, which of states, initial or final, should determine the
nonlinear terms. These issues will be discussed separately. Of
highest priority are the questions about time scales of the
processes that contribute to the transition amplitudes
(\ref{eq:E7.2}) as well as about stability of the uniform
distribution of the invariant density.

\sectionn{Summary}

~{\bf 1. The method.~~}  The most intriguing discovery of this work
is that Dirac field endows spacetime with a matter-induced affine
geometry (MIAG), which is {\it fully determined by  a real matter}.
This is possible solely because the Dirac field satisfies equations
of motion. Then, and only then, the geometry is independent of a
particular coordinate background. Possibly, this result can look
strange for mathematicians.  But it should not surprise physicists,
who know very well that nothing in spacetime can be measured without
localized material objects. So far, the method of MIAG determined
the shape of a solitary localized object as spherical dynamically
and with no conjectures. The problem of signals still has to be
worked out.

~{\bf 2. The results.~~}  The author's conjecture
\cite{Localization} that there exists a   generic mechanism of the
Dirac field  auto-localization into finite-sized positively charged
Dirac particles is rigorously confirmed. The explicit solution
representing such a particle is  found. It possesses the following
properties,\\ (i) ~~A solitary Dirac field waveform in free space
can be stable with  respect to the interaction with an external
electromagnetic field   $A_\mu$ only if this waveform is formed
solely by outward polarized components. The solution that represents
such a waveform has {\it negative energy} $E=-m$. \\ (ii)~~An
apparently complementary inward-polarized solution with negative
charge has {\it positive energy} $E=+m$. It cannot be stable as a
strongly localized object; its instability is due to the
indispensable \textquotedblleft charged currents\textquotedblright
$\Theta$ and $\Phi$. They oscillate twice faster than stationary
Dirac field, $\Theta \pm i\Phi \propto e^{\pm 2iE\tau}$. The
corresponding tetrad  components $A_{[1]},A_{[2]}$ of the vector
potential affect only the inward polarized waveform, thus making it
unstable. This \textquotedblleft motion\textquotedblright is
confined to within the spheres of a constant radius within a
localized object\footnote{This motion cannot be interpreted as an
oscillation of a {\it mean coordinate}     -- the famous
Schr$\ddot{o}$dinger {\it Zitterbewegung}.}. Similar oscillations
also show up in the theory of the Compton scattering as the
$t$-channel transitions of electron into the negative energy states.
These transitions are responsible for the classical part of the
Compton cross-section (Thompson scattering) .\\ (iii)~~The
difference in degree and the time duration of the localization
obviously makes the localized charges of opposite sign unequivocally
different particles. The correlation between the signs of electric
charge, shape and polarization explains the interdependence between
the discrete $C$- and $P$-transformations as a natural property of
the simplest localized waveforms. While $C$ qualitatively stands for
the charge conjugation, $P$ is not an abstract reflection symmetry
in a flat space; it stands for the interchange of {\it inward} and
{\it outward}. In a sense, these two discrete transformations do not
exist separately; in this sense, $CP$ is a physical symmetry between
the corresponding processes \footnote{This is in contrast with the
view of Dirac  field as the representation of the Lorentz group. In
that framework, the Poincar$\acute{\rm e}$ invariance is presumed,
and     all states can be obtained from a single state by a sequence
of the Lorentz transformations.}.

~{\bf 3. The prospects.~~}      Our major perception of vacuum is
absence of localized matter. This means that in the vacuum ${\cal
R}$ is constant, e.g., ${\cal R}=1$. Since Dirac equation is a
hyperbolic system, the Dirac field must experience refraction
towards domains where ${\cal R}>1$,  amplifying ${\cal R}$ even
more, which  resembles a well-known nonlinear effect of
self-focusing. The opposite trend     must be observed in domains
where ${\cal R}<1$; the Dirac waves tend to escape them. This idea
can be phrased more precisely as: {\it Identification of the sign of
$~log{\cal R}$ with the sign of electric charge leads to a dynamic
picture of an empirically known charge-asymmetric world in which
stable  positively charged elementary Dirac objects are highly
localized (and presumably heavy), while negatively charged objects
tend to be poorly localized (and presumably light).} This mechanism
of localization is generic and points to the picture that stunningly
resembles the today's world. It must be worked out in greater
details with the prospect that the issue of cosmological charge
asymmetry, first   addressed long ago  by A.D.Sakharov
\cite{Sakharov1}, as well as the currently observed positron excess
\cite{Serpico}, could be better understood.

Meanwhile, to validate our approach in cosmological context, two
major questions must be answered, \\ (i)~~~What (if anything) can
trigger a spontaneous creation of a proton alone (without an
antiproton)? This is the most  formidable problem.  \\(ii)~~~ Let a
$p{\bar p}$ pair is created in an energetic process and the
antiproton is thoroughly isolated from a normal matter (except for
the cosmic background radiation). Will it live infinitely long? If
not, then how will it decay? This question does not seem
unbearable\footnote{Some theories of Grand Unification  predict
proton's decay with a lifetime greater than the currently estimated
age of Universe. From our perspective, only antiproton can be
unstable.} and can be solved by methods developed in this one and
previous author's papers (work in progress).

{\color{myaqua}

\vskip 6mm
\noindent\Large\bf Acknowledgments}

\vskip 3mm

{ \fontfamily{times}\selectfont
\noindent
I am  indebted  M.E. Osinovsky for his advice on subtle issues of spinor
analysis and differential geometry and for critically reading the manuscript.
This work is supported by the Rapid Research, Inc.
{\color{myaqua}
\newpage
\appendix

\sectionn{Notation and algebraic conventions.}
\label{app:appA}
\renewcommand{\theequation}{A.\arabic{equation}}
\setcounter{equation}{0}
\color{black}
All observables associated with the Dirac field are  bilinear
forms built with the aid of {\it Hermitian} Dirac matrices
$\alpha^i=(\alpha^i)^+$ and $\beta=\beta^+$, which satisfy the
commutation relations
\begin{equation}\label{eq:A.1}
\alpha^a \beta \alpha^b  +  \alpha^b \beta \alpha^a= 2\beta
\eta^{ab}~,~~~~\alpha^a \beta  + \beta \alpha^a=0,~~~ \beta^2=1,
\end{equation}
Throughout this paper, the Dirac matrices associated with a tetrad
$h^\mu_a \in \mathbb{R}^4 $ are numeric and are chosen in the
spinor representation,
\begin{eqnarray}\label{eq:A.2}
\alpha^0=\left( \begin{array}{c c} {\bf 1} &  0 \\ 0& {\bf
1}\end{array}\right),~~~~~~~\alpha^i=\left( \begin{array}{c c}
\tau_i  &  0 \\ 0& -\tau_i\end{array}\right),
~~~~~~~\sigma^i=\left(
\begin{array}{c c} \tau_i  &  0 \\ 0&
\tau_i\end{array}\right)\nonumber\\ \rho_1=\left( \begin{array}{c
c} 0  & {\bf 1}  \\
{\bf 1}& 0 \end{array}\right),~~
\rho_2=\left( \begin{array}{c c} 0 &  -i\cdot{\bf 1}  \\
i\cdot{\bf 1} & 0 \end{array} \right),~~
\rho_3=\left( \begin{array}{c c} {\bf 1}  &  0  \\
0& -{\bf 1} \end{array} \right).
\end{eqnarray}
where   $\tau_i$ are the $2\times 2$ Pauli matrices.

If the Dirac spinor is written down in terms of modules and phases
of its components,
\begin{eqnarray}\label{eq:A.3}
\psi=\left[ \begin{array}{c} u_R \exp\big({i\phi_R^u}\big)
\\d_R \exp\big({i\phi_R^d}\big)\\u_L \exp\big({i\phi_L^u}\big)
\\d_L \exp\big({i\phi_L^d}\big)\end{array}\right],
\end{eqnarray}
then, with the Dirac matrices (\ref{eq:A.2}), the scalars and the
four Dirac currents have the following components,
\begin{eqnarray}\label{eq:A.4}
j^a=\left[ \begin{array}{c} u_L^2 +d_L^2+ u_R^2 +d_R^2
\\ 2u_L d_L\cos\bigg(\phi_L^u-\phi_L^d\bigg)- 2u_R d_R\cos\bigg(\phi_R^u-\phi_R^d\bigg)
\\ -2u_L d_L\sin\bigg(\phi_L^u-\phi_L^d\bigg)+ 2u_R d_R\sin\bigg(\phi_R^u-\phi_R^d\bigg)
\\ u_L^2 -d_L^2- u_R^2 +d_R^2 \end{array}\right], \nonumber\\
{\cal J}^a=\left[ \begin{array}{c} u_L^2 +d_L^2-
u_R^2-d_R^2 \\ 2u_L d_L\cos\bigg(\phi_L^u-\phi_L^d\bigg)+ 2u_R
d_R\cos\bigg(\phi_R^u-\phi_R^d\bigg) \\ -2u_L d_L\sin\bigg(\phi_L^u-\phi_L^d\bigg)-
2u_R d_R\sin\bigg(\phi_R^u-\phi_R^d\bigg)
\\ u_L^2 -d_L^2+ u_R^2 -d_R^2 \end{array}\right],
\end{eqnarray}
\begin{eqnarray}\label{eq:A.5}
\Theta^a=\left[ \begin{array}{c}- 2u_L
d_R\cos\bigg(\phi_L^u+\phi_R^d\bigg)+ 2d_L u_R\cos\bigg(\phi_R^u+\phi_L^d\bigg) \\
2u_L u_R \cos\bigg(\phi_L^u+\phi_R^u\bigg)- 2d_L d_R\cos\bigg(\phi_R^d+\phi_L^d\bigg)
\\ -2u_L u_R\sin\bigg(\phi_L^u+\phi_R^u\bigg)- 2d_L d_R\sin\bigg(\phi_R^d+\phi_L^d\bigg)
\\ - 2u_L d_R\cos\bigg(\phi_L^u+\phi_R^d\bigg)-
2d_L u_R\cos\bigg(\phi_R^u+\phi_L^d\bigg) \end{array}\right], \nonumber\\
\Phi^a=\left[ \begin{array}{c}- 2u_L d_R\sin\bigg(\phi_L^u+\phi_R^d\bigg)+ 2d_L
u_R\sin\bigg(\phi_R^u+\phi_L^d\bigg) \\  2u_L u_R \sin\bigg(\phi_L^u+\phi_R^u\bigg)-
2d_L d_R\sin\bigg(\phi_R^d+\phi_L^d\bigg) \\ 2u_L
u_R\cos\bigg(\phi_L^u+\phi_R^u\bigg)+ 2d_L d_R\cos\bigg(\phi_R^d+\phi_L^d\bigg)
\\ - 2u_L d_R\sin\bigg(\phi_L^u+\phi_R^d\bigg)-
2d_L u_R\sin\bigg(\phi_R^u+\phi_L^d\bigg) \end{array}\right],
\end{eqnarray}
\begin{eqnarray}\label{eq:A.6}
{\cal S}\!+\!i{\cal P}=2\bigg(u_R u_L
e^{i(\phi_L^u-\phi_R^u)}\! + d_R d_L
e^{i(\phi_L^d-\phi_R^d)}\bigg)={\cal R}e^{i{\cal Y}} ,~,\nonumber\\
{\cal R}^2\!=\!4[u_R^2 u_L^2+d_R^2 d_L^2+2u_R u_L d_R d_L
\cos(\phi_L^u\!-\phi_R^u\!-\phi_L^d\!+\phi_R^d)].~~~
\end{eqnarray}

\end{document}